\documentclass[
  journal=largetwo,
  manuscript=article-type,
  year=2024,
  volume=37,
]{cup-journal}

\usepackage{amsfonts}
\usepackage{amssymb}
\usepackage[nopatch]{microtype}
\usepackage[table]{xcolor}
\usepackage{booktabs}
\usepackage{physics}
\usepackage{caption}
\usepackage{hyperref}
\usepackage{biblatex}
\usepackage{xcolor}
\usepackage{soul}
\usepackage{graphics}
\usepackage{float}
\usepackage{amsmath}
\usepackage{pdfpages}
\usepackage{aas-macros}

\hypersetup{
	colorlinks=true,
	linkcolor=blue,
	filecolor=blue,
	urlcolor=blue,
	citecolor=blue
}

\title{On bursty star formation during cosmological reionization - how does it influence the baryon mass content of dark matter halos?}

\author{Anand Menon}
\affiliation{International Centre for Radio Astronomy Research, The University of Western Australia, 35 Stirling Highway, Crawley, Western Australia 6009, Australia}

\author{Chris Power}
\affiliation{International Centre for Radio Astronomy Research, The University of Western Australia, 35 Stirling Highway, Crawley, Western Australia 6009, Australia}
\alsoaffiliation{ARC Centre of Excellence for All Sky Astrophysics in 3 Dimensions (ASTRO 3D)}

\email[Anand Menon]{anand-jm@hotmail.com}


\begin{document}

\begin{abstract}
The baryon mass content (i.e. stellar and gas mass) of dark matter halos in the early Universe depends on both global factors - e.g. ionising ultraviolet (UV) radiation background - and local factors - e.g. star formation efficiency and assembly history. We use a lightweight semi-analytical model to investigate how both local and global factors impact the halo baryon mass content at redshifts of $z\geq 5$. Our model incorporates a time delay between when stars form and when they produce feedback of $0\leq t^d/\text{Myr} \leq 30$, which can drive bursts of star formation, and a mass and redshift dependent UV background, which captures the influence of cosmological reionization on gas accretion onto halos. We use statistically representative halo assembly histories and assume that the cosmological gas accretion rate is proportional to the halo mass accretion rate. Delayed ($t^d$>0) feedback leads to oscillations in gas mass with cosmic time, behaviour that cannot be captured with instantaneous feedback ($t^d$=0). Highly efficient star formation drives stronger oscillations, while strong feedback impacts when oscillations occur; in contrast, inefficient star formation and weak feedback produce similar long-term behaviour to that observed in instantaneous feedback models. If the delayed feedback timescale is too long, a halo retains its gas reservoir but the feedback suppresses star formation. Our model predicts that lower mass systems (halo masses $m_\text{h} \leq 10^7 \text{M}_\odot$) at $z \leq 10$ should be strongly gas deficient ($m_\text{g}\rightarrow 0$), whereas higher mass systems retain their gas reservoirs because they are sufficiently massive to continue accreting gas through cosmological reionization. Interestingly, in higher mass halos, the median $m_\star/(m_\star+m_\text{g}) \simeq 0.01-0.05$, but is a factor of 3-5 smaller when feedback is delayed. Our model does not include seed supermassive black hole feedback, which is necessary to explain massive quenched galaxies in the early Universe. 
\end{abstract}

\section{Introduction}
\label{sec:introduction}
    \par Recent observations with the JWST have revealed a potential overabundance of massive galaxies ($M_{\star}\sim 10^{10.5-11} {\rm M}_{\odot}$) in the high redshift Universe ($z\gtrsim 7$) when compared to the predictions of theoretical models of galaxy formation \citep[e.g.][]{Finkelstein2023,Labbe2023}. This has prompted questions about the validity of the standard cosmological model and the early growth of dark matter halos (e.g. \citealt[][]{boylan-kolchin2023}, but see \citealt{KraghJespersen2024}), as well as questions about our understanding of the interplay between gas reservoirs, star formation, and feedback at early times \citep[e.g.][]{Dekel2023,Boyett2024}. Whether or not there is an actual overabundance of massive galaxies when compared to theoretical model predictions \citep[see discussion in, e.g., ][]{Finkelstein2023_2}, the questions these observations have prompted highlight the fundamental time limit imposed by the age of the Universe at early cosmic times when assessing galaxy properties, given the rate at which the highest mass dark matter halos can grow and consequently the stellar masses of the galaxies that they contain.

    \par A similar time limit exists at lower halo masses. It has been understood for some time that the presence of photo-ionizing backgrounds should inhibit the formation of galaxies in low mass dark matter halos \citep[e.g.][]{Efstathiou1992,thoul1996}. For example, the presence of an ultraviolet (UV) background will photo-ionize warm diffuse gas in the inter-galactic medium (IGM), preventing its accretion onto low mass dark matter halos while also preventing gas already present in halos from cooling\citep[e.g.][]{Efstathiou1992}. Such a UV background emerged during cosmological reionization driven by high mass stars \citep[e.g.][]{Barkana2001,Wyithe2003}, leading to the suppression of galaxy formation in lower mass halos \citep[e.g.][]{Gnedin2000,Wyithe2006,Okamoto2008,Kravtsov2022}. Therefore, for lower mass halos to host galaxies, we might argue that there is a time limit imposed by the requirement that they form their stars prior to cosmological reionization, at $z\gtrsim 6$. 

    \par In this paper, we investigate the consequences of the time limit imposed by cosmological reionization for the baryon content - the stellar mass, $m_\ast$, and gas mass, $m_\text{g}$ - of dark matter halos at high redshifts. In particular, we examine the interplay between the UV background-driven suppression of gas accretion onto lower dark matter halos, stellar feedback driven by star formation in these halos, and their baryon mass content - $m_\ast$ and $m_\text{g}$. Galaxies embedded within low-mass halos will experience episodes of supernova-driven feedback that deplete their reservoir of star forming gas via powerful winds \citep[e.g.][]{Dekel1986,Efstathiou2000}, and prevent further stellar mass assembly and gas accretion. It's interesting to note that the main sequence lifetimes of the high mass stars ($M \geq 8 {\rm M}_{\odot}$) that result in supernovae are relatively short - of order $10^7$ yrs - which, at $z\gtrsim 6$, is an appreciable fraction - approximately 10\% - of a dark matter halo's dynamical time. This means that a common assumption of galaxy formation models - that the time between star formation and supernova feedback is short and be treated as instantaneous - should break down in the high redshift Universe. Instead, feedback is delayed, leading to episodic or bursty star formation \citep[e.g.][]{Scalo1986,FaucherGiguere2018,Orr2019,Furlanetto2022,Pallottini2023A&A,Shen2023,Sun2023ApJ}.
    
    \par \textcite{FaucherGiguere2018} modelled such bursty star formation and showed that it can arise when the effective equilibrium between the self-gravity of gas in a galaxy and stellar feedback following star formation cannot be maintained. This should be commonplace in galaxies at high redshifts and in lower mass galaxies across cosmic time. Indeed, there is observational evidence for bursty star formation in galaxies at high redshifts \citep[e.g.][]{Faisst2019,Looser2023,Strait2023}. \textcite{Faisst2019} concluded from a statistical sample of $z\!\sim$4.5 galaxies that the significant scatter in UV and H$\alpha$ luminosities and star formation rates implied episodes of bursty star formation during the prior $50$ Myr. Drawing on JWST/NIRSpec data, \textcite{Strait2023} found star formation is in the process of being shut down in a $z=5.2$ galaxy, while \textcite{Looser2023} found star formation has been quenched in a $z=7.3$ galaxy; both are consistent with galaxy formation model predictions that incorporate bursty star formation \citep{Dome2023}. Similarly, there is observational evidence for bursty star formation in dwarf galaxies \citep[e.g.][]{Weisz2012,Emami2019}, based on the distributions of UV and H$_{\alpha}$ that probe their star formation histories. Such bursty star formation arises naturally in hydrodynamical galaxy formation simulations that can track resolved star formation and feedback \citep[e.g.][]{Hopkins2023}, especially in lower mass galaxies \citep[e.g.][]{Onorbe2015,Muratov2015,Sparre2017} and during the epoch probed by JWST \citep[e.g.][]{Pallottini2023A&A,Shen2023,Sun2023ApJ}.   
    
    \par Using a lightweight semi-analytical model that we have written, we examine how the combination of UV suppression of gas accretion from cosmological scales and bursty star formation driven by delayed feedback influence the baryon content of dark matter halos at high redshifts. We model the assembly histories of these halos using Monte Carlo merger trees that have been calibrated against cosmological $N$-body simulations \citep[cf.][]{Parkinson2008} and therefore capture plausible variations in their growth over cosmic time. We assume that cosmological gas accretion tracks halo growth, modulated by the presence of the UV background, following the approach of \citet[][]{Kravtsov2022}. We parameterise star formation and feedback efficiencies, allowing for delayed feedback following the approach of \citet{Furlanetto2022}. In this way, we can assess how local factors (star formation efficiency, onset and strength of stellar feedback, variation in dark matter halo assembly history) and global factors (onset of a UV ionizing background and the suppression of cosmological gas accretion) influence the baryon content of halos at $z\geq 5$.
    
    \par In the following sections, we describe the main features of our model (\S~\ref{sec:model}); we present our results for the baryon content of halos as a function of cosmic time and halo mass (\S~\ref{sec:results}), showing how they are sensitive to the assumptions of our model (instantaneous versus delayed feedback, absence or presence of UV background), model parameters (e.g. star formation efficiency, delayed feedback time interval), and variations in dark matter halo assembly history; and we summarise our key findings and their implications for our understanding of galaxy formation and evolution in the low-mass halo regime (\S~\ref{sec:conclusions}). We use the cosmological parameter values $\Omega_\text{b}=0.0484$, $\Omega_\text{m}=0.308$, $\Omega_\Lambda=0.692$, $h=0.678$, $\sigma_8=0.815$, and $n_\text{s}=0.968$, which are consistent with the results obtained by the \citet{Planck2018}.

\section{Theoretical Model}
\label{sec:model}
    We have written a lightweight semi-analytical model\footnote{This python code can be made available by contacting the corresponding author.} for studying the growth of gas and stellar mass in dark matter halos, which is motivated by the philosophy of equilibrium models set out in \textcite{Dave2012}. We track the mass assembly of dark matter halos and the corresponding cosmological accretion of gas from the IGM; the accumulation of a cold gas reservoir at the centre of the halo; star formation; and the expulsion of gas via stellar feedback (supernovae-driven winds). In particular, we follow \textcite{Furlanetto2022} and introduce a delay between when stars form and when they produce feedback, and we adopt a time-dependent suppression of cosmological gas accretion that reflects the growth of a UV ionizing background, following \textcite{Kravtsov2022}. 
    
    \par Note that we do not account for the growth of the seeds of super-massive black holes and the feedback they produce. We expect this to be important in more massive galaxies - star formation will be preferred at the expense of black hole growth in lower mass galaxies because black hole growth is limited to occur on a Salpeter timescale \citep[see, e.g.][]{Nayakshin2009MNRAS,Bourne2016}. This difference in timescales on which stellar and black hole driven feedback acts \citep[e.g.][]{Power2011} should give rise to more complex bursty star formation histories in more massive galaxies; we will investigate this in a future paper.

    \subsection{Evolutionary Equations}
	We track the time rate of change of gas ($\dot{m}_{\text{g}}$), star formation ($\dot{m}_{\star}$), and supernova-driven wind mass loss ($\dot{m}_{w}$) at time $t$ via a set of coupled differential equations:
	\begin{align}
		\dot{m}_{\text{g}}(t)&=\dot{m}_\text{c,g}(t)-\dot{m}_{\star}(t)-\dot{m}_\text{w}(t),\label{1.1}\\
		\dot{m}_{\star}(t)&=\epsilon_\text{sf}\frac{m_\text{g}(t)}{\tau_\text{sf}},\label{1.2}\\
		\dot{m}_\text{w}(t)&=\eta_\text{fb}\dot{m}_\star(t-t^\text{d})\label{1.3}.
	\end{align}
    It is straightforward to convert between $t$ and redshift $z$ for our adopted cosmology, which is important for evaluating the contribution of the UV background. The various parameters are now described in the following subsections.

    \subsubsection{Growth of the Gas Reservoir}
    \par Equation~\ref{1.1} tracks the gas reservoir in a galaxy, accounting for the accretion rate of gas from cosmological scales ($\dot{m}_\text{c,g}$), gas that is converted to stars via $\dot{m}_{\star}$, and gas that is lost via winds ($\dot{m}_w$). As is commonly assumed \citep[e.g.][]{White1991}, $\dot{m}_{\text{c,g}}$ tracks the accretion rate of the dark matter halo that hosts the galaxy, $\dot{m}_\text{h}$; 
    \begin{equation}
		\dot{m}_\text{c,g}=\epsilon_\text{in}\qty(\frac{\Omega_\text{b}}{\Omega_\text{m}})\dot{m}_\text{h}.
        \label{1.4}
	\end{equation}
	Here the quantities $\Omega_\text{b}$ and $\Omega_\text{m}$ refer to the total baryon and matter densities of the Universe, respectively, and the assumption is that the halo accretes a baryon mass equivalent to the cosmic fraction ($\Omega_\text{b}/\Omega_\text{m}$) per unit dark matter accreted. We also use a pre-factor $0 \leq \epsilon_\text{in} \leq 1$, such that $\epsilon_\text{in}=1$ corresponds to the accretion of the cosmic baryon fraction and $\epsilon_\text{in}=0$ corresponds to complete suppression of baryon accretion. We discuss this further in the context of UV suppression of cosmological accretion below.

    \subsubsection{Growth of Stellar Mass}   
	\par Equation~\ref{1.2},
    \begin{equation}    
        \dot{m}_{\star}=\epsilon_\text{sf}\frac{m_{g}}{\tau_\text{sf}}, \nonumber  
    \end{equation}    
     links the star formation rate to the mass of the gas reservoir, $m_\text{g}$, a star formation timescale, $\tau_\text{sf}$, and an efficiency factor, $\epsilon_\text{sf}$. We assume that $\tau_\text{sf}$ is proportional to the dynamical time, 
     \begin{equation}
         \tau_\text{sf} \propto t_\text{dyn}=\frac{R}{\sigma}
     \end{equation}
     where $R$ and $\sigma$ are the characteristic radius and velocity dispersion of the system respectively. If we adopt the dynamical time of the halo as characterising this timescale, we expect that 
     \begin{equation}
         t_\text{dyn}=\left(\frac{4}{\Delta_\text{vir}}\right)^{1/2} \frac{1}{H(t)} \simeq \frac{0.15}{H(t)}
     \end{equation}
     where $\Delta_\text{vir} \simeq 178$ is the virial overdensity of the halo and
 	 \begin{equation}
	 	H(t)=H_\text{0}\sqrt{\Omega_\text{m}(1+z(t))^3+\Omega_\Lambda}.
	 \end{equation}
     is the Hubble parameter at cosmic time, $t\equiv~t(z)$. We assume that      
     \begin{equation}
         \tau_\text{sf} = 0.15 \frac{f_\text{sf}}{H(z)}
     \end{equation}
     and expect that $f_\text{sf} \gtrsim 1$. 
     We take $\epsilon_\text{sf}=0.015$ as our fiducial value, following \textcite{Furlanetto2022} and motivated by \citet{Murray2011} and \citet{Leroy2017}, but we allow this parameter to vary between $\epsilon_\text{sf}=0.0015$ (low efficiency) and $\epsilon_\text{sf}=0.1$ (high efficiency) to gauge its influence on our results. 

    \subsubsection{Supernovae-Driven Wind Mass Loss}
    \label{ssec:wind}
    \par Equation~\ref{1.3},
    \begin{equation}
 		\dot{m}_\text{w}=\eta_\textbf{fb}\dot{m}_\star(t-t^\text{d}),\nonumber
    \end{equation}
    links the rate of mass loss driven by supernovae to $\dot{m}_\star$ at some time $t-t^\text{d}$, where $t^\text{d}$ is the delayed feedback timescale. If $t^\text{d}=0$, then this mass loss rate is driven by the instantaneous star formation rate; however, if $t^\text{d}>0$, then it depends on star formation at some earlier time and corresponds to delayed feedback. We adopt $t^\text{d}$=0.015 Gyr (15 Myr) as our fiducial value, but have explored values of $0 \leq t^\text{d} \leq 0.03$ Gyr. 
    
    \par We follow \textcite{Furlanetto2017} in defining our choice of feedback efficiency parameter; they model momentum-regulated supernova feedback and assume that supernovae accelerate the wind to the escape velocity of the halo. This gives, 
	\begin{equation}		\eta_\textbf{fb}=\epsilon_\text{fb}\pi_\text{p}\qty(\frac{10^{11.5}M_{\odot}}{m_\text{h}})^{1/3}\qty(\frac{9}{1+z})^{1/2},\label{1.10}
	\end{equation}
	where $\epsilon_\text{fb}$ parameterises feedback efficiency and corresponds to the momentum injected by the supernovae driving the wind, and $\pi_{\text{p}}$ is the total amount of momentum from each supernova. We follow \textcite{Furlanetto2022} in adopting $\epsilon_\text{fb}=5$ as our fiducial value, but we also consider values in the range $2\leq \epsilon_\text{fb}\leq 7$. Following \cite{Furlanetto2017}, we assume $\pi_{\text{p}}=2\times 10^{33}$ g cm $\text{s}^{-2}$.

    \subsubsection{UV Suppression of Cosmological Gas Accretion}
	\par We follow the approach of \textcite{Kravtsov2022} by noting that the gas mass within a dark matter halo may be a function of both mass and redshift \citep[e.g.][]{Okamoto2008}, 
    \begin{equation}
        m_\text{g}(z)=f_\text{b}(m_\text{h},z)m_\text{h}(z)\label{2.0}.
    \end{equation}
    The quantity $f_\text{b}(m_\text{h},z)$ is related to $\epsilon_\text{in}$ in
    Equation~\ref{1.1}, as we make clear below. Following \textcite{Gnedin2000,Okamoto2008}, we write,
	\begin{equation}
		f_\text{b}(m_\text{h},z)=\frac{\Omega_\text{b}}{\Omega_\text{m}}s(\mu_\text{c},\omega),\label{2.1}
	\end{equation}
	where $\Omega_\text{b}/\Omega_\text{m}$ is the cosmic baryon fraction and $s(x,y)$ is defined as,
	\begin{equation}
		s(x,y)=\qty[1+(2^{\frac{y}{3}}-1)x^{-y}]^{-\frac{3}{y}}\label{2.1a}; 
	\end{equation}
	$\mu_\text{c}=m_\text{h}/M_\text{c}(z)$, where $M_\text{c}(z)$ is a characteristic mass below which $s(\mu_\text{c},\omega) \rightarrow 0$. Following \citet{Kravtsov2022}, we adopt
	\begin{equation}
		M_\text{c}=1.69\times 10^{10}\frac{\text{exp}(-0.63z)}{1+\text{exp}([z/\beta]^{\gamma})}M_{\odot},\label{2.2}
	\end{equation}
	where $\gamma=15$ and $\beta$ is given by,
    \begin{equation}		\beta=z_\text{rei}\qty\Bigg[\text{ln}\qty\Big(1.82\cross10^3\text{exp}(-0.63z_\text{rei})-1)]^{-1/\gamma};
	\end{equation}
	here $z_\text{rei}$ is the redshift of reionization. We adopt as our fiducial value $z_\text{rei}=7$ but have checked the sensitivity of our results to this choice.

    \par If we consider the time rate of change of Equations~\ref{2.0} and~\ref{2.1}, we see that
    \begin{align}
        \dot{m}_\text{g}(t)=\dot{f}_\text{b}m_\text{h}+f_\text{b}\dot{m}_\text{h}&=\left(f_\text{b}+\dot{f}_\text{b}\frac{m_\text{h}}{\dot{m}_\text{h}}\right)\dot{m}_\text{h}\nonumber\\
        &=\left(s+\dot{s}\frac{m_\text{h}}{\dot{m}_\text{h}}\right)\left(\frac{\Omega_\text{b}}{\Omega_\text{m}}\right)\dot{m}_\text{h}\nonumber,
    \end{align}
    which is equivalent to Equation~\ref{1.4}, and so we can evaluate $\epsilon_\text{in}$, given the form of Equation~\ref{2.1a}.\footnote{We note that \textcite{Kravtsov2022} explored the general case in which gas accretion is suppressed by the presence of a UV background ($\epsilon_\text{UV}$), the presence of a hot halo ($\epsilon_\text{hot}$), and preventive feedback ($\epsilon_\text{prev}$), such that $\epsilon_\text{in}=\epsilon_\text{UV}\epsilon_\text{hot}\epsilon_\text{prev}$. We argue that suppression by hot halos and by preventive feedback can be neglected at high redshifts (i.e. $\epsilon_\text{hot}\epsilon_\text{prev}=1$, and so we consider $\epsilon_\text{in}=\epsilon_\text{UV}$.} We write,
         \begin{align}                       
            \epsilon_\text{in}=&\text{max}\bigg(0,s\qty(\mu_\text{c},\omega)\bigg[(1+X)-\nonumber\\
            &2\epsilon(z,\gamma)\frac{M_\text{h}}{\dot{M}_\text{h}}X(1+z)H(z)\bigg]\bigg),\label{2.5}
        \end{align}
	where 
	\begin{align*}
		\epsilon(z,\gamma)&=\frac{0.63}{1+e^{(z/\beta)^\gamma}}+\frac{\gamma z^{\gamma-1}}{\beta^\gamma}\frac{e^{(z/\beta)^\gamma}}{(1+e^{(z/\beta)^\gamma})^2},\\
		X&=\frac{3c_\omega M_\omega}{1+c_\omega M_\omega},\quad c_\omega=2^{\omega/3}-1,\\ M_\omega&=\qty(\frac{M_\text{c}(z)}{M_\text{h}})^\omega,\quad\omega=2.
	\end{align*}
	This has the effect of suppressing gas accretion at halo masses below the threshold mass $M_\text{c}(z)$ at $z<z_\text{rei}$, with $M_\text{c}(z)$ increasing with decreasing $z$ (i.e. accretion is suppressed onto progressively higher mass halos). For redshifts $z>z_\text{rei}$, there is no suppression of gas accretion. We refer the interested reader to Figure 1 of \textcite{Kravtsov2022} for an illustration of this behaviour.

    \subsection{Modelling Dark Matter Halo Growth}
    We have used the algorithm of \textcite{Parkinson2008} to generate Monte Carlo dark matter halo merger trees based on Extended Press-Schechter theory \citep[e.g.][]{Bond1991,Lacey1993}. This has been calibrated to provide predicted halo masses as a function of cosmic time that are consistent with merger trees derived from the Millennium Simulation \citep[cf.][]{Springel2005}. For each of the 11 halo mass bins equally spaced (linearly) in the interval $10^6 \text{M}_{\odot} \leq m_\text{h} \leq 10^{11} \text{M}_{\odot}$, we sample 100 realisations following mass assembly histories in the redshift range $5 \leq z \leq 25$ equally spaced in the logarithm of the expansion factor, $a=1/(1+z)$. This provides us with $m_\text{h}$ as a function of $z$; we calculate $\dot{m}_\text{h}$ as a function of $t(z)$ by constructing a smooth spline interpolant, which we differentiate numerically.

    \begin{figure}[!h]
		\centering
		\includegraphics[width=\linewidth]{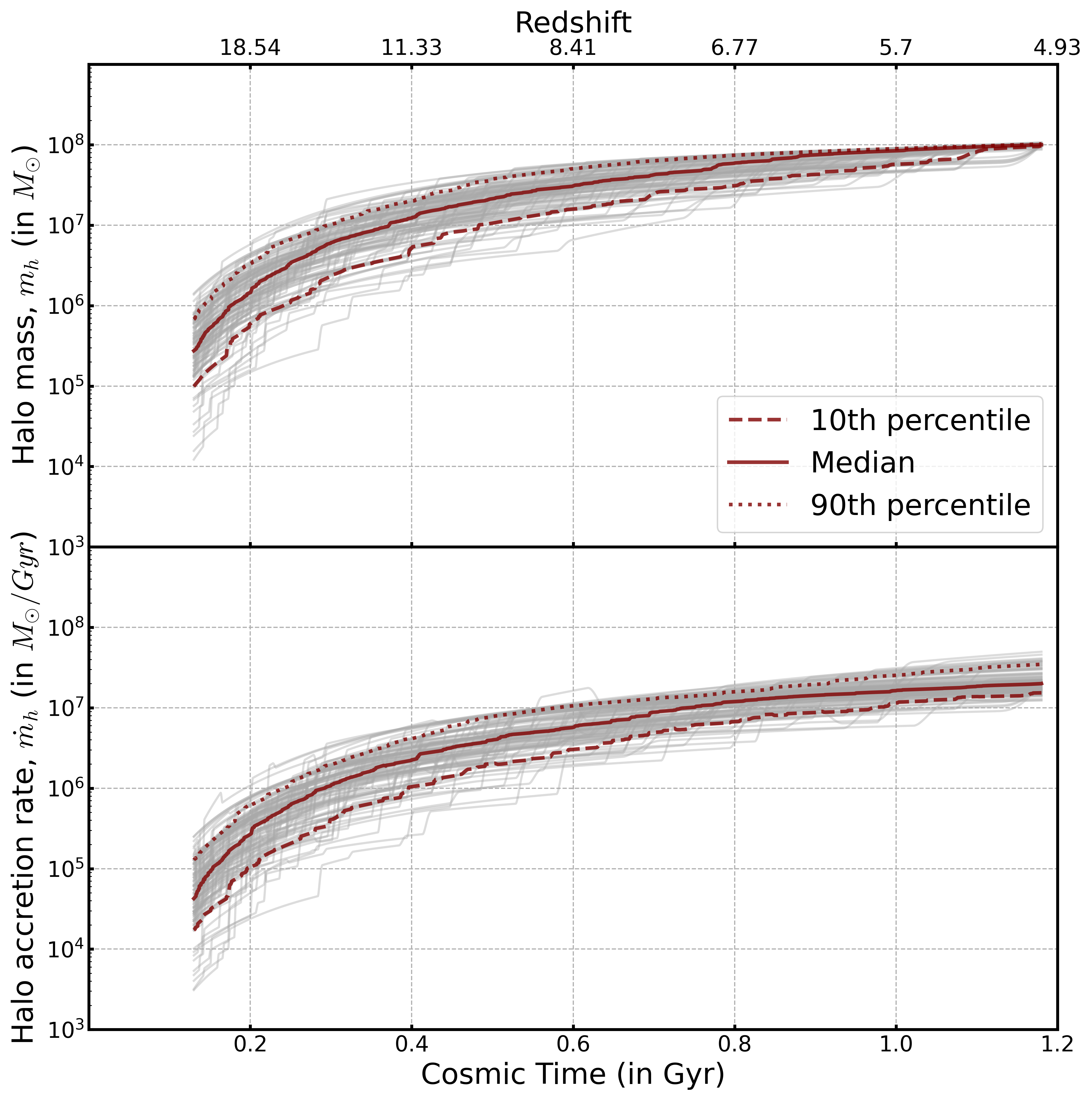}
		\caption{Mass assembly history of a sample of 100 dark matter halos with a halo mass of $m_\text{h}=10^8\text{M}_{\odot}$ at $z$=5 generated using the \textcite{Parkinson2008} Monte Carlo merger tree algorithm. The upper and lower panels show $m_\text{h}$ (in $\text{M}_{\odot}$) and $\dot{m}_\text{h}$ (in $\text{M}_{\odot}/{\text{Gyr}}$) against cosmic time (in Gyrs; lower horizontal axis) and redshift (upper horizontal axis); red solid and dotted curves indicate the median, 10$^\text{th}$, and 90$^\text{th}$ percentiles of the distributions at a given time.}
		\label{fig:HaloAssemblyHistory}
	\end{figure}
 
    \par In Figure~\ref{fig:HaloAssemblyHistory}, we show the mass assembly histories of a sample of dark matter halos that have masses of $m_\text{h}=10^8\text{M}_{\odot}$ at $z=5$. Each individual halo is represented by a light grey curve; the median, 10$^\text{th}$, and 90$^\text{th}$ percentiles are indicated by the red solid and dotted curves. The upper panel shows the variation of $m_\text{h}$ with cosmic time (lower horizontal axis) and redshift (upper horizontal axis), starting at $z$=25, while the lower panel shows the corresponding values of $\dot{m}_\text{h}$. This shows that our Monte Carlo merger trees produce a diversity of assembly histories and that halos that have the same value of $m_\text{h}$ at $z$=5 can differ by up to an order of magnitude in $m_\text{h}$ and $\dot{m}_\text{h}$ at earlier times. This allows us to explore how such variation in $\dot{m}_\text{h}$ and the corresponding influence on $\dot{m}_\text{c,g}$ and $\dot{m}_\star$ affect the halos' baryon mass content.
 
    \section{Results}
    \subsection{Bursty star formation \& UV suppression of accretion}
    \label{sec:results}
    \begin{figure}[!h]
		\centering
		\includegraphics[width=\linewidth]{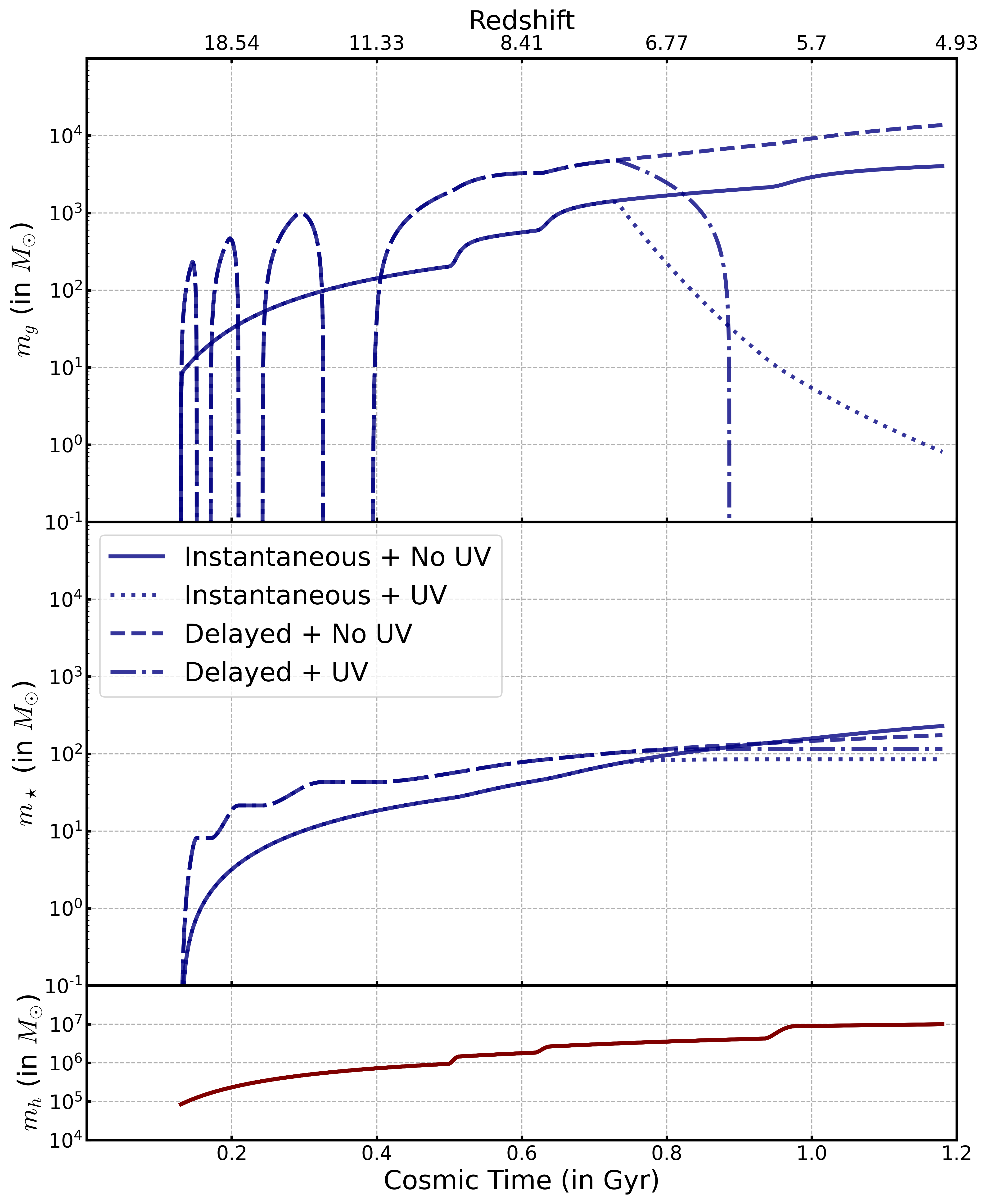}
		\caption{Baryon mass assembly history of an example halo with $m_\text{h}=10^7\text{M}_{\odot}$ at $z$=5. The upper and middle panels show $m_\text{g}$ and $m_\star$ (in $\text{M}_{\odot}$) against cosmic time (in Gyrs); solid (dotted) curves correspond to instantaneous star formation and feedback without (with) UV suppression of accretion (instantaneous + No UV, instantaneous + UV), while dashed (dot-dashed) curves correspond to bursty star formation from delayed feedback without (with) UV suppression of accretion (delayed + No UV, delayed + UV). For comparison, we show also the growth of $m_\text{h}$ with cosmic time in the lower panel.}
		\label{fig:BaryonAssemblyHistory}
	\end{figure}
    We begin by considering the growth of gas mass (upper panel), stellar mass (middle panel), and halo mass (lower panel; for comparison) in Figure~\ref{fig:BaryonAssemblyHistory}, focusing on an example halo with $m_\text{h}=10^7\text{M}_{\odot}$ at $z=5$, drawn from the sample of 100 halos in this mass bin. We choose this halo mass bin because it marks the transition between low-mass halos that can be efficiently quenched via delayed feedback alone and high-mass halos that continue to accrete and form stars after the onset of UV suppression; we show examples of low- and high-mass systems' assembly histories in \ref{sec:appendix_massdependence}. To illustrate the effects of bursty star formation and UV suppression on the growth of gas and stellar mass as a function of cosmic time, we consider the cases of instantaneous star formation and feedback and no UV suppression of accretion (hereafter instantaneous + No UV; solid curves); instantaneous star formation feedback and UV suppression of gas accretion with $z_\text{rei}=7$ (hereafter instantaneous + UV; dotted curves); bursty star formation from delayed feedback and no UV suppression of accretion (hereafter delayed + No UV; dashed curves); and bursty star formation and feedback and  UV suppression of accretion with $z_\text{rei}=7$ (hereafter delayed + UV; dot-dashed curves). We adopt our fiducial parameters of $\epsilon_\text{sf}=0.015$ and $\epsilon_\text{fb}=5$.

    \par In the instantaneous + No UV case, we find that both $m_\text{g}$ and $m_\star$ show an initial steep rise at early times ($\leq 0.5$ Gyr), and continue to grow but at a slower at later times (up to 2 Gyr). There are noticeable increases in $m_\text{g}$ at approximately 0.5 and 0.6 Gyr; these correlate with increases in $m_\text{h}$ evident in the lower panel, which translate into increases in $m_\text{g}$ given our assumption that $\dot{m}_\text{c,g}$ tracks $\dot{m}_\text{h}$. In contrast, for instantaneous + UV, we find that $m_\text{g}$ and $m_\star$ show a similar initial steep rise at early times but the effect of the UV background shutting off cosmological accretion is evident - $m_\text{g}$ peaks after 0.75 Gyr and then declines with increasing time, while $m_\star$ plateaus. Because $m_\text{g}$ is linked to $\dot{m}_\text{h}$ and the magnitude of UV suppression is both halo mass and redshift dependent, we see that $m_\text{g}$ shows an initial sharp decline but plateaus after 1.5 Gyr. For the delayed + No UV case, we see the characteristic strong oscillations in $m_\text{g}$ - and to a lesser extent in $m_\star$ - evident during the initial 0.5 Gyr, but as the halo grows, these oscillations damp away, and both quantities are indistinguishable from the case with instantaneous star formation and feedback. Similarly, for the delayed + UV case, we see these oscillations in $m_\text{g}$ and $m_\star$ repeated.

   \subsection{Halo mass assembly history}
   \begin{figure}[!h]
		\centering
		\includegraphics[width=\linewidth]{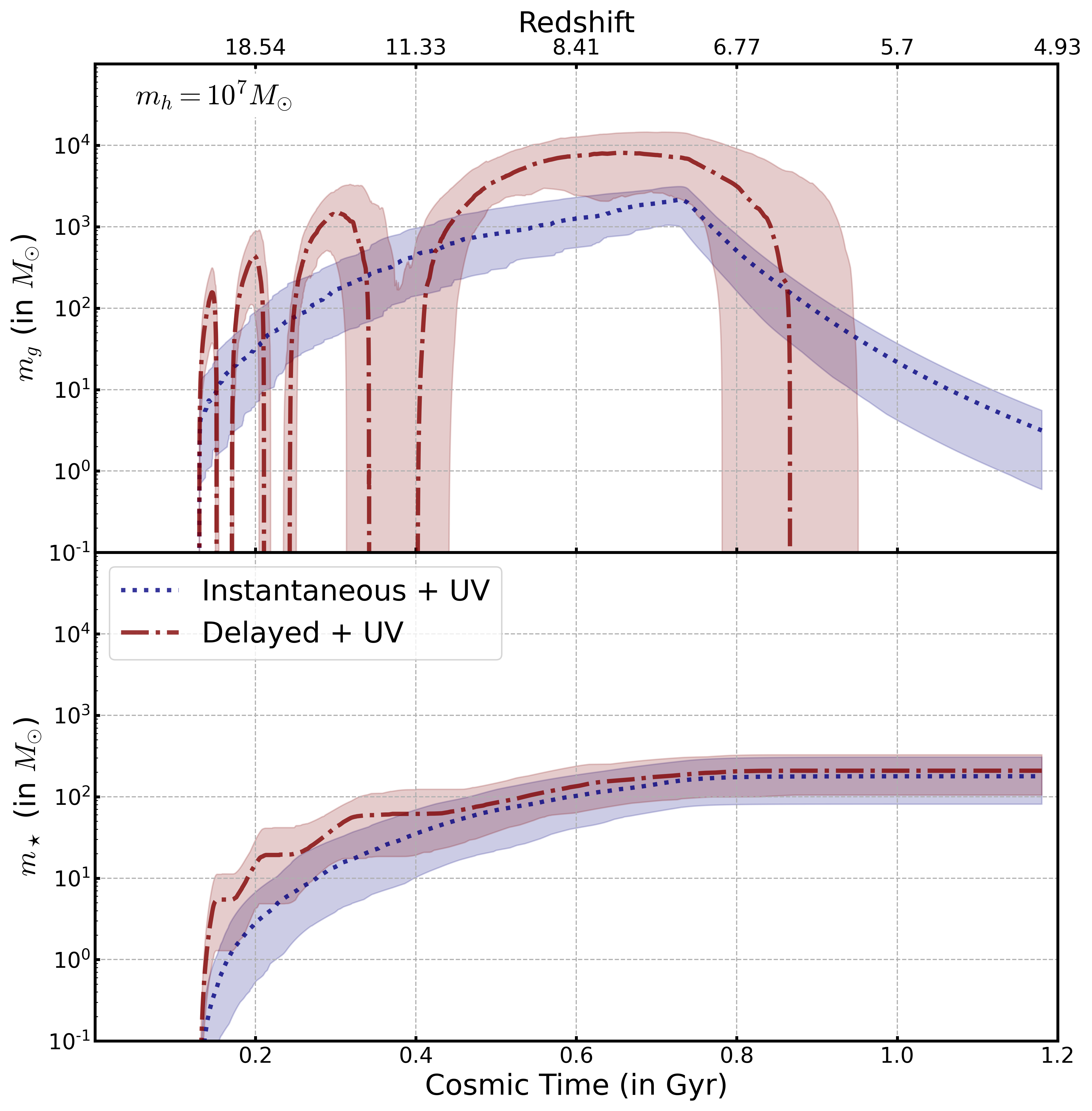}
		\caption{\textbf{Influence of halo mass assembly history}: Baryon mass assembly history of a sample of 100 halos with $m_\text{h}=10^7\text{M}_{\odot}$ at $z$=5. The upper and lower panels show the median values of $m_\text{g}$ and $m_\star$ (in $\text{M}_{\odot}$) against cosmic time (in Gyrs); dotted curves correspond to the instantaneous + UV case, while dot-dashed curves correspond to the delayed + UV case. The coloured bands indicate the range of the 10$^\text{th}$ and 90$^\text{th}$ percentiles.}
		\label{fig:BaryonAssemblyHistoryPopulation}
	\end{figure}   
   \par In Figure~\ref{fig:BaryonAssemblyHistoryPopulation}, we show the variation in $m_\text{g}$ and $m_\star$ with cosmic time for a sample of 100 halos, each with $m_\text{h}=10^7\text{M}_{\odot}$ at $z=5$. We now focus on models with the UV suppression of accretion of gas from the IGM. The dotted (dot-dashed) curves correspond to the median values of $m_\text{g}$ (upper panel) and $m_\star$ (lower) for the instantaneous + UV (delayed +UV) case, while the coloured bands indicate the range of the 10$^\text{th}$ and 90$^\text{th}$ percentiles. This shows how variations in the assembly history of the underlying dark matter halo, whose growth rate $\dot{m}_\text{h}$ governs the growth rate of the gas mass and consequently the stellar mass. The variations introduced by the diversity of halo assembly histories correspond to approximately $0.5-1$ dex in both $m_\text{g}$ and $m_\star$. The variations in $m_\text{g}$ are larger for the delayed + UV case, as we might expect, while we we see that the variations in $m_\star$ are similar in magnitude for both the instantaneous + UV and delayed + UV cases ($\sim 0.5$ dex at early times, $\sim 1$ dex at later times).
   
    \subsection{Efficiency of star formation, $\epsilon_\text{sf}$, and feedback, $\epsilon_\text{fb}$}
   \begin{figure}[!h]
		\centering
		\includegraphics[width=\linewidth]{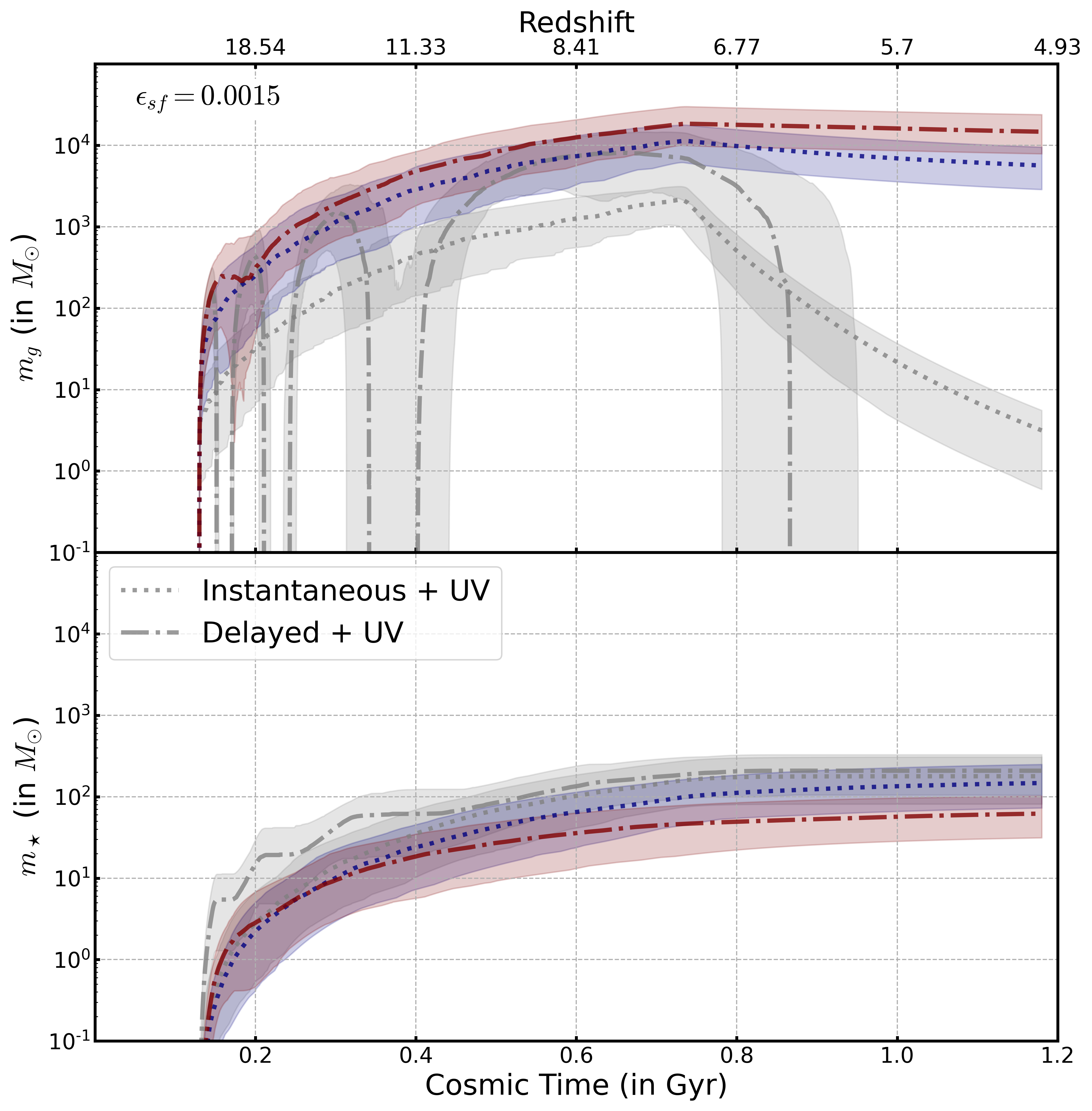}
  		\includegraphics[width=\linewidth]{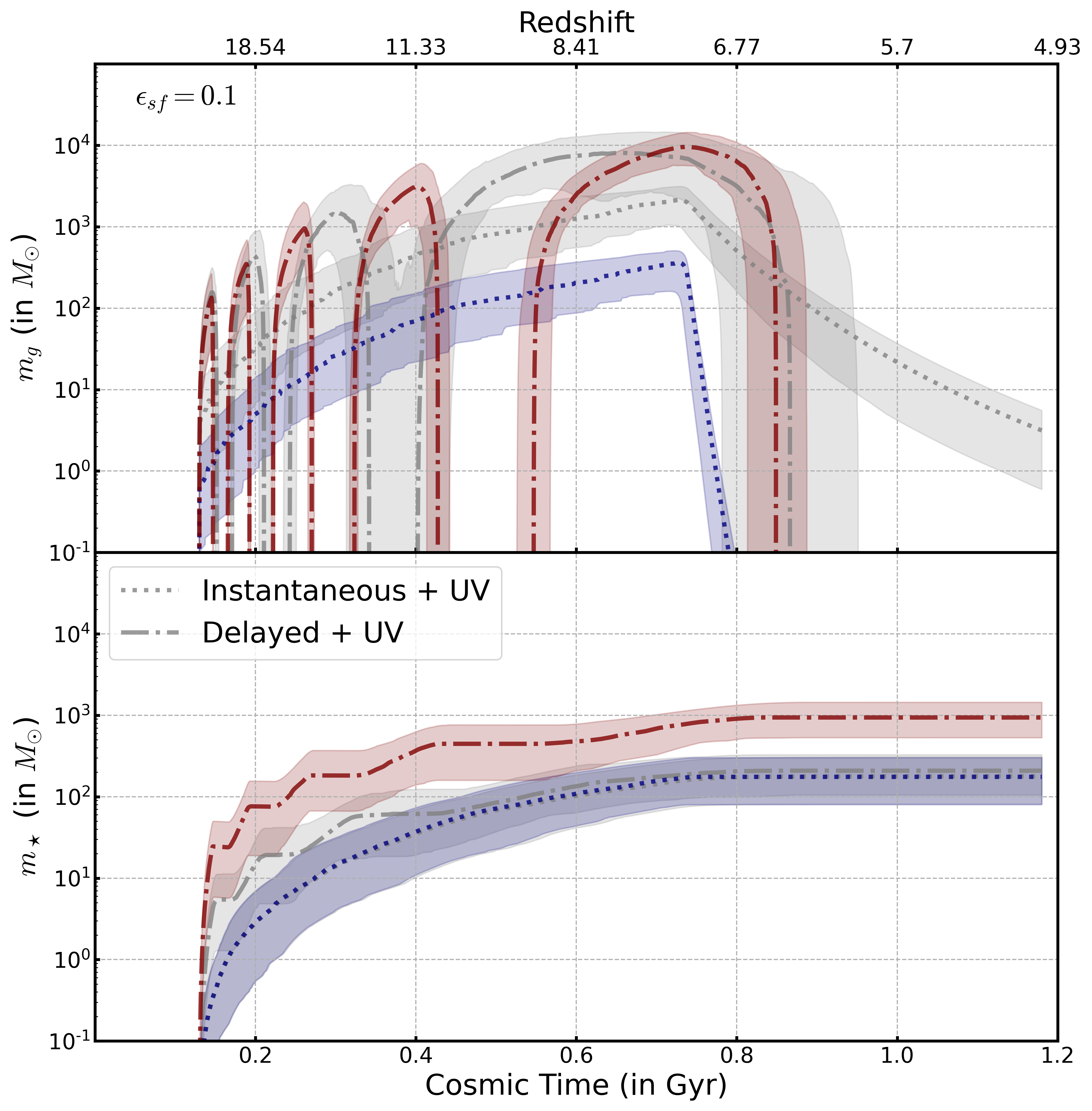}
		\caption{\textbf{Influence of star formation efficiency}: Baryon mass assembly history of a sample of 100 halos with $m_\text{h}=10^7\text{M}_{\odot}$ at $z$=5. The upper and lower panels show the median values of $m_\text{g}$ and $m_\star$ (in $\text{M}_{\odot}$) against cosmic time (in Gyrs) for $\epsilon_\text{sf}$=0.0015 (inefficient) and $\epsilon_\text{sf}$=0.1 (efficient) respectively. Dotted curves correspond to the instantaneous + UV case, while dot-dashed curves correspond to the delayed + UV case. The coloured bands indicate the range of the 10$^\text{th}$ and 90$^\text{th}$ percentiles. The greyed bands and curves correspond to the counterpart cases with the fiducial value of $\epsilon_\text{sf}$=0.015.}
		\label{fig:BaryonAssemblyHistorySFEfficiency}
	\end{figure}
    
\begin{figure}[!h]
		\centering
		\includegraphics[width=\linewidth]{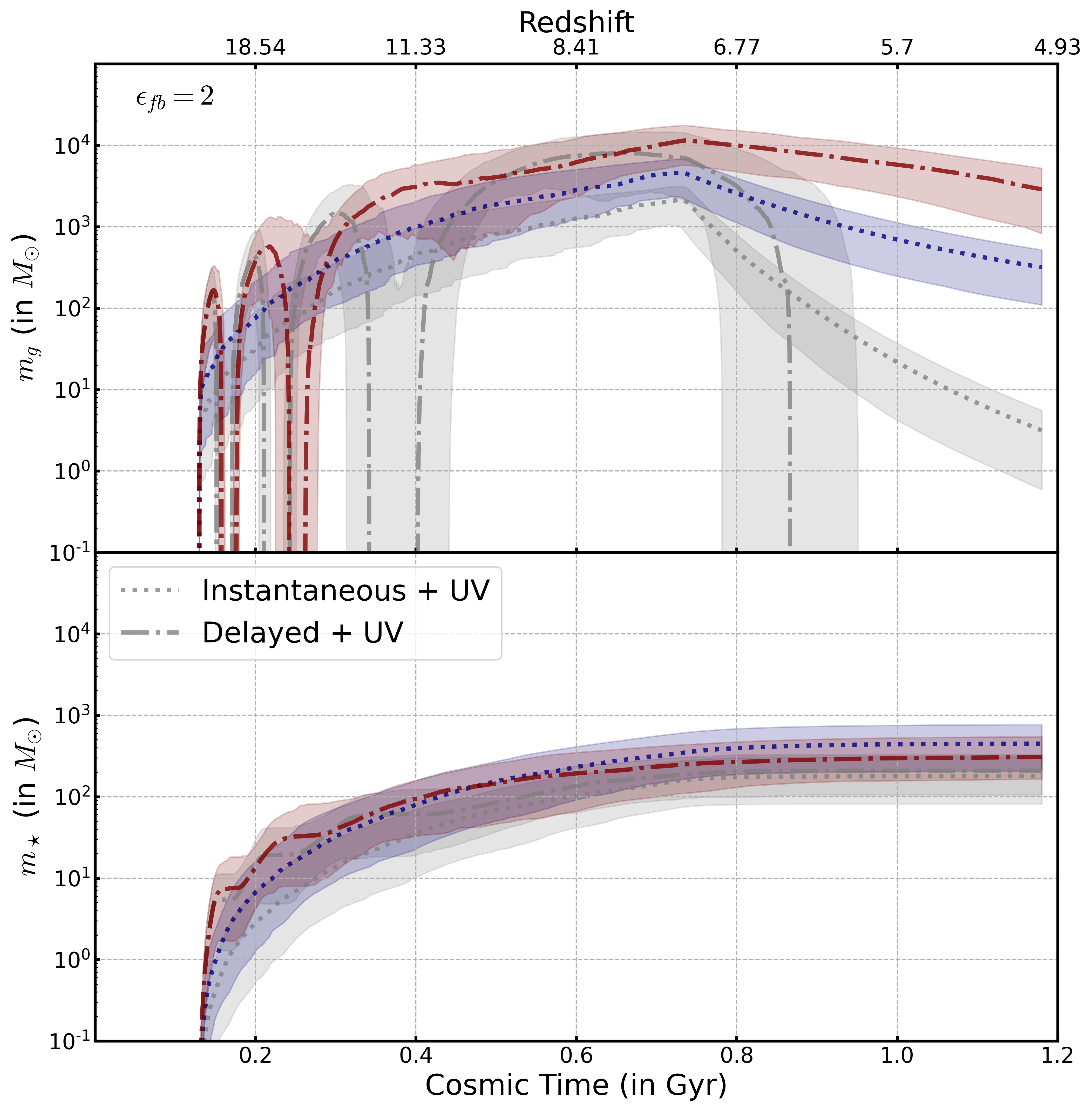}
  		\includegraphics[width=\linewidth]{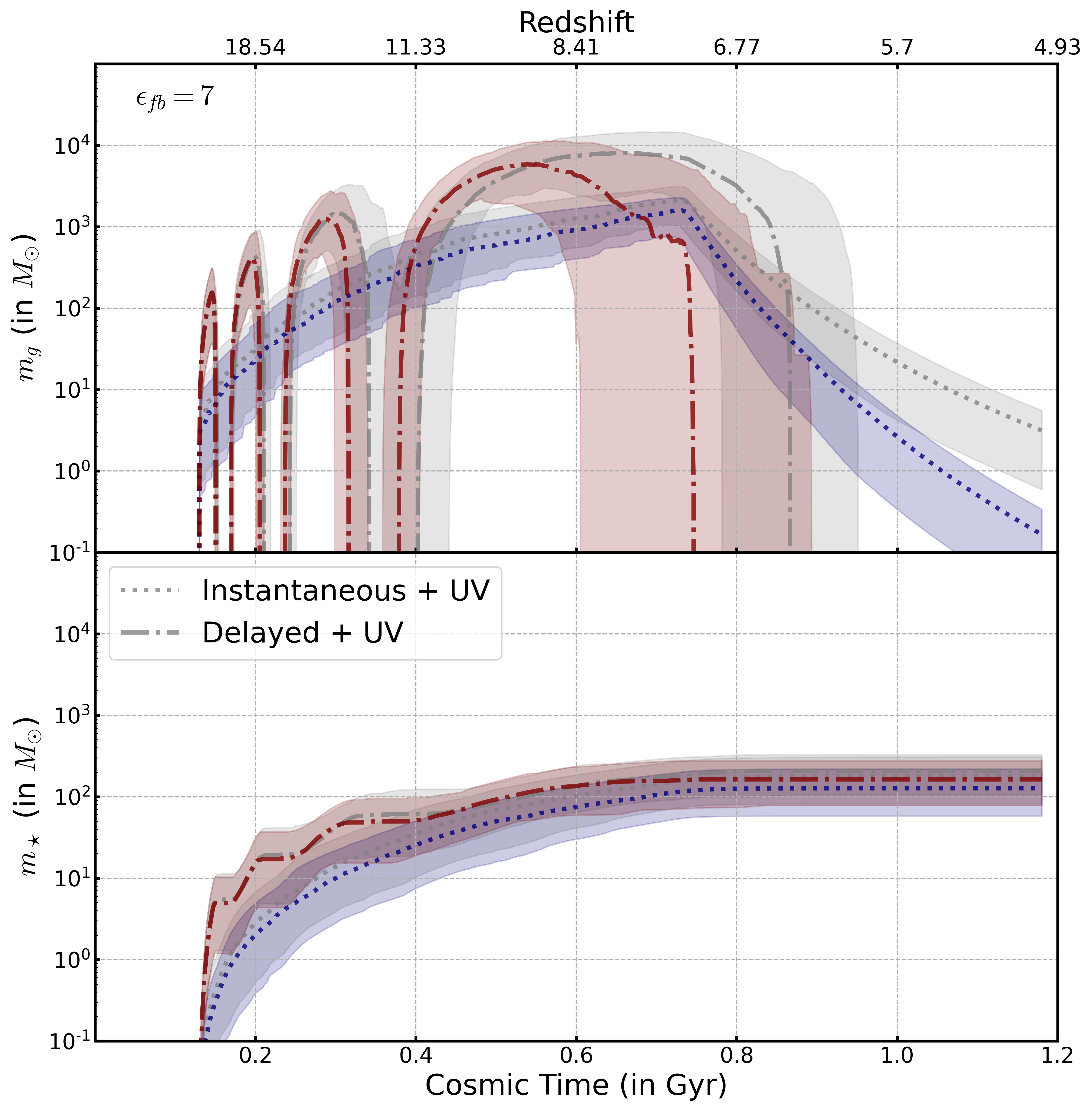}
		\caption{\textbf{Influence of feedback efficiency}: Baryon mass assembly history of a sample of 100 halos with $m_\text{h}=10^7\text{M}_{\odot}$ at $z$=5. The upper and lower panels show the median values of $m_\text{g}$ and $m_\star$ (in $\text{M}_{\odot}$) against cosmic time (in Gyrs) for $\epsilon_\text{fb}$=2 (inefficient) and $\epsilon_\text{fb}$=7 (efficient) respectively. Dotted curves correspond to the instantaneous + UV case, while dot-dashed curves correspond to the delayed + UV case. The coloured bands indicate the range of the 10$^\text{th}$ and 90$^\text{th}$ percentiles. The greyed bands and curves correspond to the counterpart cases with the fiducial value of $\epsilon_\text{fb}$=5.}
		\label{fig:BaryonAssemblyHistoryFBEfficiency}
	\end{figure}

    We now examine how our choices of star formation and feedback efficiency affect the growth of gas mass (upper panel) and stellar mass (lower panel) in Figures~\ref{fig:BaryonAssemblyHistorySFEfficiency} and~\ref{fig:BaryonAssemblyHistoryFBEfficiency}. As in Figure~\ref{fig:BaryonAssemblyHistoryPopulation}, we focus on a sample of 100 halos with $m_\text{h}=10^7\text{M}_{\odot}$ at $z=5$, but for cases of low and high star formation and feedback efficiencies, $\epsilon_\text{sf}$ and $\epsilon_\text{fb}$, respectively. We keep $z_\text{rei}=7$ fixed and look at the cases of low (high) efficiency in the upper (lower) panels, with reference curves corresponding to our fiducial values plotted in light grey ($\epsilon_\text{sf}$=0.015 and $\epsilon_\text{fb}$=5). 
    
    \par As before, we focus on the cases of instantaneous + UV (dotted curves) and delayed + UV (dot-dashed curves). By reducing (increasing) the star formation efficiency, we expect a reduction in the number of stars to form over a fixed interval to be lower (higher) than in the fiducial case; however, the coupling of star formation to feedback means that a higher star formation efficiency should boost the strength of the feedback, which, in the case of delayed feedback, should lead to stronger oscillations in the ${m}_\text{g}$ and ${m}_\star$ \citep[see, e.g.,][]{Furlanetto2022}. We anticipate that stronger (weaker) feedback efficiency should impact $\dot{m}_\text{c,g}$ and consequently $m_\text{g}$, but it's likely the that the difference with respect to the fiducial run will be more pronounced for weaker feedback - it may evacuate the halo of gas more quickly, but the accretion rate onto the halo is unchanged.

    \par In Figure~\ref{fig:BaryonAssemblyHistorySFEfficiency}, we fix feedback efficiency at $\epsilon_\text{fb}$=5 and look at star formation efficiencies of $\epsilon_\text{sf}$=0.0015 (low efficiency, upper panel) and $\epsilon_\text{sf}$=0.1 (high efficiency, lower panel). As expected, in the case of higher star formation efficiency, we see evidence for enhanced star formation efficiency driving enhanced feedback. Consider the instantaneous + UV case. At a given time, we see that $m_\text{g}$ is smaller by approximately 0.5 dex compared to the fiducial run up to the point at which the effect of UV suppression kicks in, following which $m_\text{g}$ rapidly declines as its converted to stars; the evolution of $m_\star$ is relatively unchanged from the fiducial case. In contrast, the delayed + UV case shows some interesting changes compared to the fiducial case; the initial evolution of $m_\text{g}$ is similar but from approximately 0.2 Gyr it experiences 3 longer periods in which $m_\text{g}\rightarrow 0$, while we also note that the typical system forms more stars at a given time, and $m_\star$ is $\sim 1$ dex higher after 1.2 Gyr. For the low star formation efficiency case, we see that the differences between the two models are dramatically reduced. In particular, the strong oscillations apparent in the fiducial delayed + UV case are very weak - only a dip in $m_\text{g}$ is evident at 0.2 Gyr - and the evolution in $m_\text{g}$ and $m_\star$ are very similar, with the delayed + UV case having higher values of $m_\text{g}$ by $\sim 0.1$ dex and lower values of $m_\star$ by $\sim 0.1$ dex, when compared to the instantaneous + UV case.

    \par In Figure~\ref{fig:BaryonAssemblyHistoryFBEfficiency}, we fix star formation efficiency at $\epsilon_\text{sf}$=0.015 and look at feedback efficiencies of $\epsilon_\text{fb}$=2 (low efficiency, upper panel) and $\epsilon_\text{fb}$=7 (high efficiency, lower panel). As anticipated, stronger feedback acts to accelerate when the median halo loses its gas mass (i.e. $m_\text{g}\rightarrow 0$), especially following the onset of UV suppression, and the spread in times when this occurs is larger than in the fiducial model (0.3 Gyr compared to 0.2 Gyr in the fiducial case); however, it has little effect on the evolution of the stellar mass. We find that weaker feedback has a dramatic effect on the evolution of $m_\text{g}$ in the delayed +UV model; the initial oscillations seen in the fiducial case are present, but after 0.3 Gyrs $m_\text{g}$ grows up to the onset of UV suppression, before showing a slow decline; the total gas mass in this model is larger than in the instantaneous feedback case after the initial oscillations have damped away. The behaviour of the instantaneous +UV models in both the strong and weak feedback cases track that in the fiducial case, with the key difference being in the rate of decline of $m_\text{g}$ after the onset of UV suppression - the halo retains more (less) gas mass at a given time when feedback is strong (weak). 
    
     \subsection{Delayed Feedback Timescale, $t^\text{d}$}
\begin{figure}[!h]
		\centering
		\includegraphics[width=\linewidth]{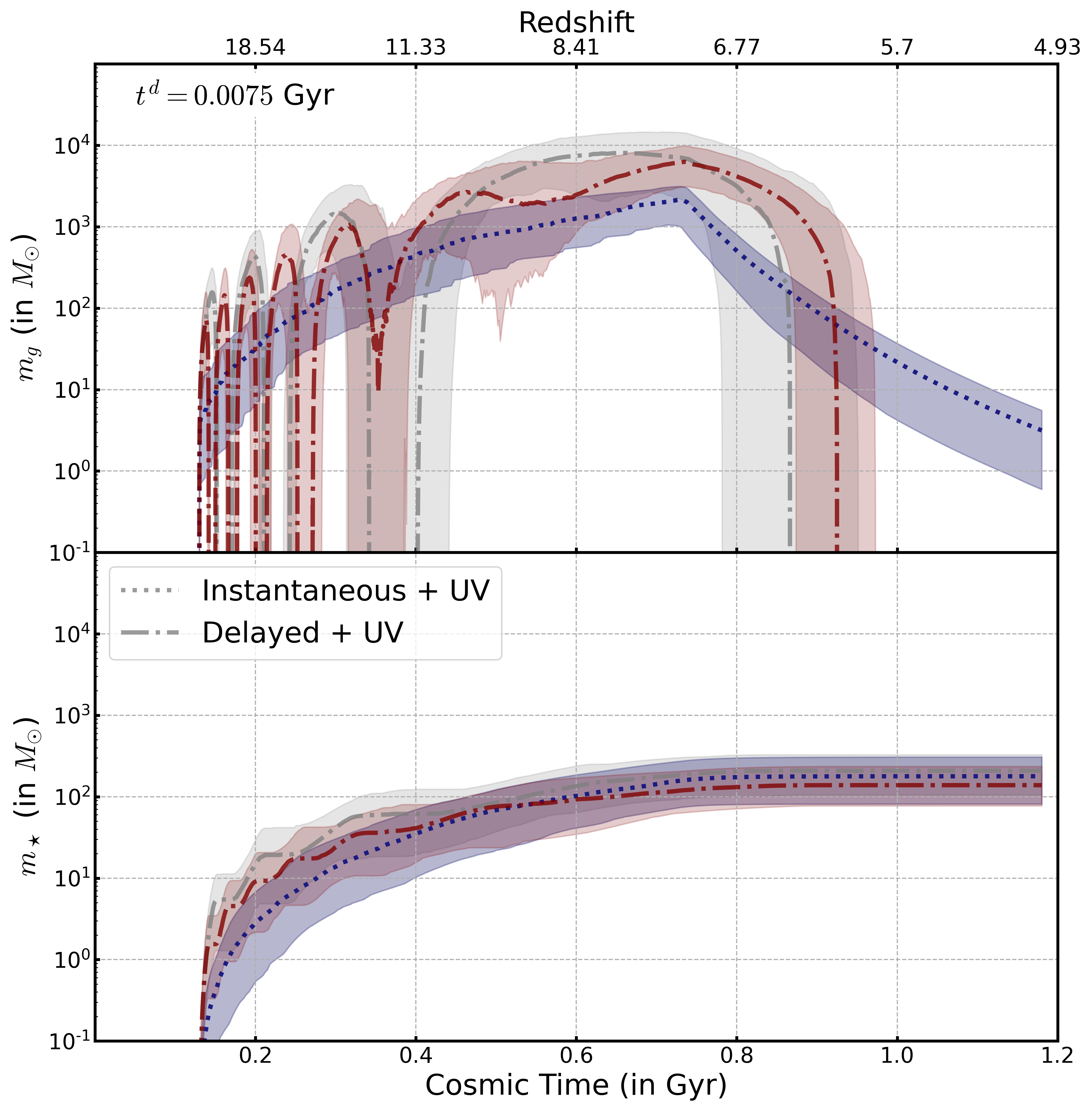}
		\includegraphics[width=\linewidth]{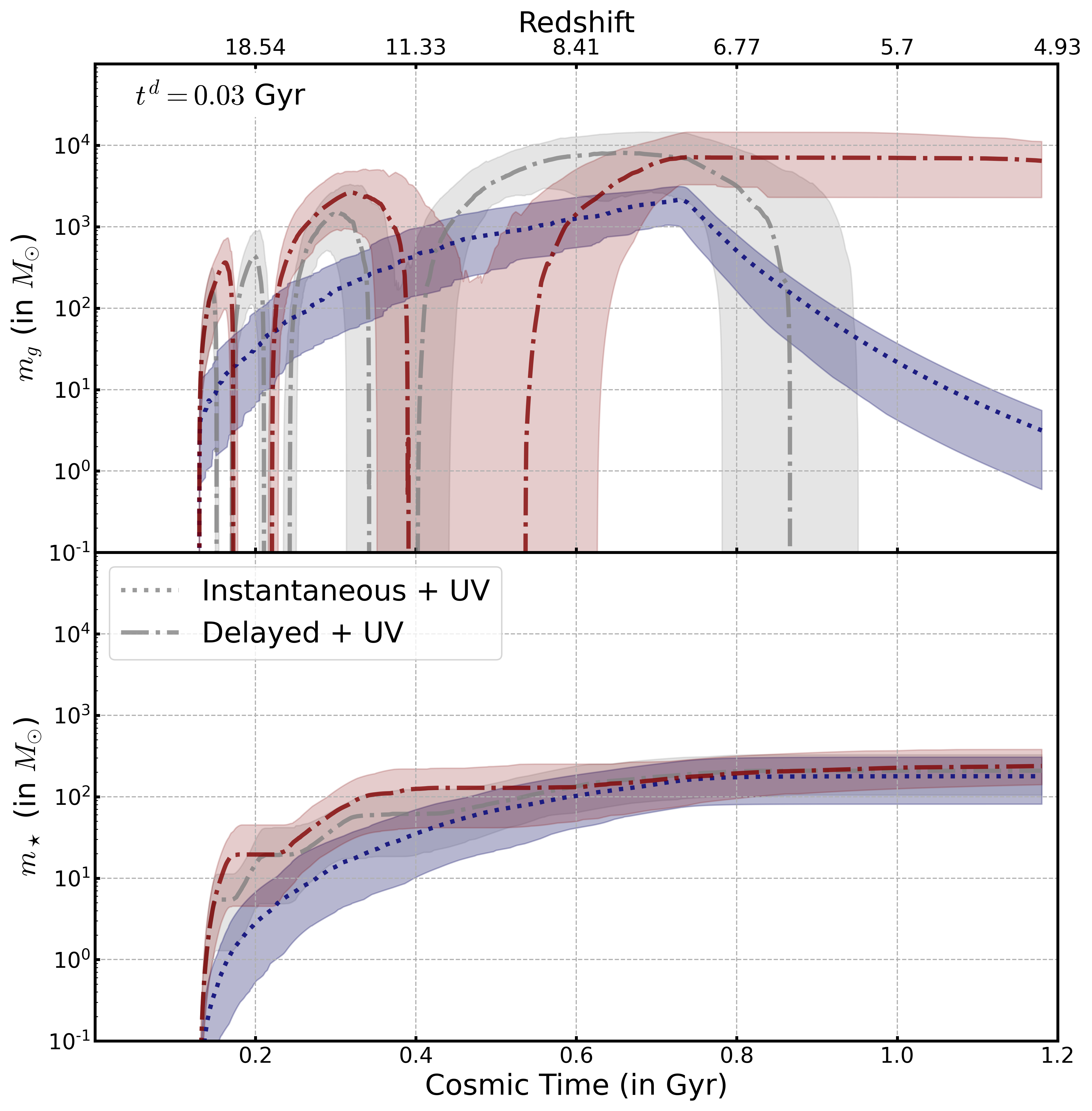}
  
		\caption{\textbf{Influence of delayed feedback timescale}: Baryon mass assembly history of a sample of 100 halos with $m_\text{h}=10^7\text{M}_{\odot}$ at $z$=5. The upper and lower panels show the median values of $m_\text{g}$ and $m_\star$ (in $\text{M}_{\odot}$) against cosmic time (in Gyrs) for $t^\text{d}$=0.0075 Gyr and $t^\text{d}$=0.03 Gyr respectively. Dotted curves correspond to the instantaneous + UV case, while dot-dashed curves correspond to the delayed + UV case. The coloured bands indicate the range of the 10$^\text{th}$ and 90$^\text{th}$ percentiles. The greyed bands and curves correspond to the counterpart cases with the fiducial value of $t^\text{d}$=0.015 Gyr.}
		\label{fig:BaryonAssemblyHistoryFBTimescale}
	\end{figure}

     In Figure~\ref{fig:BaryonAssemblyHistoryFBTimescale} we assess how our choice of the delayed feedback timescale, 
    $t^\text{d}$, influences the growth of $m_\text{g}$ and $m_\star$ with cosmic time. We focus on a sample of 100 halos with $m_\text{h}=10^7\text{M}_{\odot}$ at $z=5$, with $z_\text{rei}=7$ fixed and our fiducial values of star formation and feedback efficiency of $\epsilon_\text{sf}$=0.015 and $\epsilon_\text{fb}$=5. According to our formulation, the instantaneous star formation and feedback model corresponds to $t^\text{d}$=0, while our fiducial delayed feedback model assumes $t^\text{d}$=0.015 Gyr; we consider two further cases, with $t^\text{d}$=0.0075 Gyr (upper panel) and 0.03 Gyr (lower panel).
    
    \par Compared to our fiducial model with $t^\text{d}$=0.015 Gyr, we see that a shorter (longer) time delay increases (decreases) the number of initial oscillations in both $m_\text{g}$ and $m_\star$, and the peak median masses are similar. There are two noticeable differences - in the lengths of the intervals when $m_\text{g}\rightarrow 0$ at early times, with $t^\text{d}$=0.03 Gyr experiencing longer intervals when $m_\text{g}\simeq 0$; and in the evolution of $m_\text{g}$ after the onset of UV suppression of gas accretion, such that $m_\text{g}$ plateaus close to its peak value in the model with the long time delay through to 1.2 Gyr, whereas $m_\text{g}\rightarrow 0$ after the onset of UV suppression as in the fiducial case, albeit with a delay of $\sim 0.1$ Gyr. These trends highlight the complex interplay between gas accretion, star formation, and delayed feedback - especially the manner in which gas mass is retained at later times when $t^\text{d}$=0.03 Gyr.

    \subsection{Redshift of Reionization, $z_\text{rei}$}
\begin{figure}[!h]
		\centering
		\includegraphics[width=\linewidth]{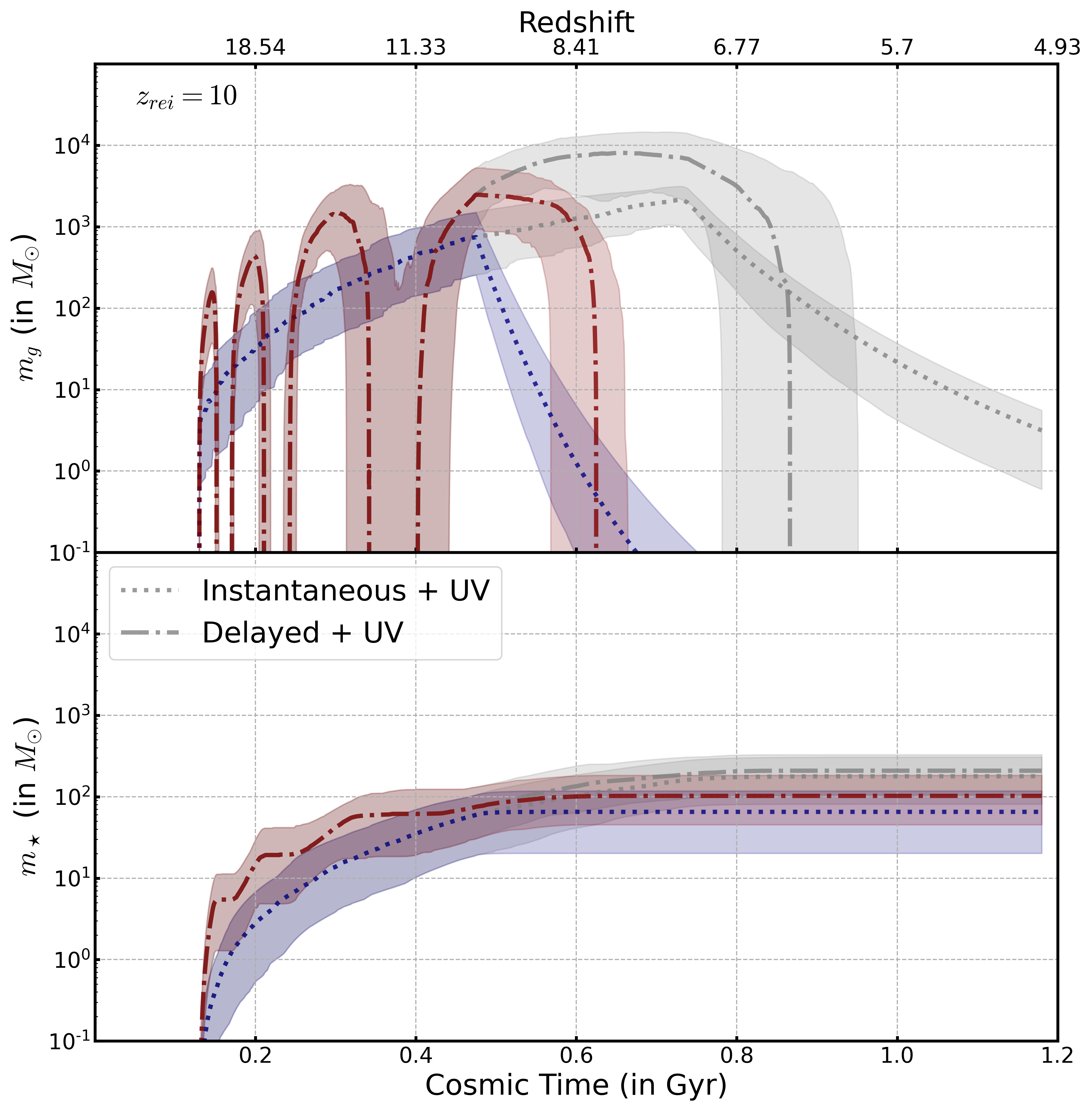}  
		\caption{\textbf{Influence of redshift of reionization}: Baryon mass assembly history of a sample of 100 halos with $m_\text{h}=10^7\text{M}_{\odot}$ at $z$=5. We show the median values of $m_\text{g}$ and $m_\star$ (in $\text{M}_{\odot}$) against cosmic time (in Gyrs) for $z_\text{rei}$=10. Dotted curves correspond to the instantaneous + UV case, while dot-dashed curves correspond to the delayed + UV case. The coloured bands indicate the range of the 10$^\text{th}$ and 90$^\text{th}$ percentiles. The greyed bands and curves correspond to the counterpart cases with the fiducial value of $z_\text{rei}$=7.}
		\label{fig:BaryonAssemblyHistoryUVSuppressionOnset}
	\end{figure}

      In Figure~\ref{fig:BaryonAssemblyHistoryUVSuppressionOnset} we assess how our choice of the redshift of reionization, $z_\text{rei}$, influences the growth of $m_\text{g}$ and $m_\star$ with cosmic time. Again we focus on a sample of 100 halos with $m_\text{h}=10^7\text{M}_{\odot}$ at $z=5$, with our fiducial values of star formation and feedback efficiency of $\epsilon_\text{sf}$=0.015, $\epsilon_\text{fb}$=5 fixed. We look at the case of $z_\text{rei}$=10, which corresponds to the time when the Universes was approximately 60\% of its age at $z$=7 (our fiducial redshift of reionization). The trends are qualitatively similar to those we see when $z_\text{rei}$=7; the peak values of $m_\text{g}$ and $m_\star$ are lower, and the rate at which $m_\text{g}\rightarrow 0$ is more rapid in the case of instantaneous feedback with UV suppression, as we would expect because of the shorter dynamical times in these systems.

    \subsection{Baryon Mass Content - Variation with Halo Mass}
    We have explored how local factors - star formation and feedback efficiencies, burstiness of star formation, variation in dark matter halo assembly histories - and global factors - the onset of UV suppression arising from cosmological reionization - influence the time evolution of the gas and stellar masses of halos. We focused on a halo mass scale - $m_\text{h}=10^7\text{M}_{\odot}$ at $z=5$ - that we argued is transitional, being sufficiently massive enough not to experience early truncation of gas accretion and star formation, but not overly massive that it's unaffected by delayed feedback and UV suppression of gas accretion from the IGM. We have this information for halos spanning a mass range $10^6 \text{M}_\odot\leq m_\text{h} \leq 10^{11}\text{M}_\odot$ at that epoch, as we show in \ref{sec:appendix_massdependence}. 
    
    \par
\begin{figure*}[!h]
		\centering
		\includegraphics[width=\linewidth]{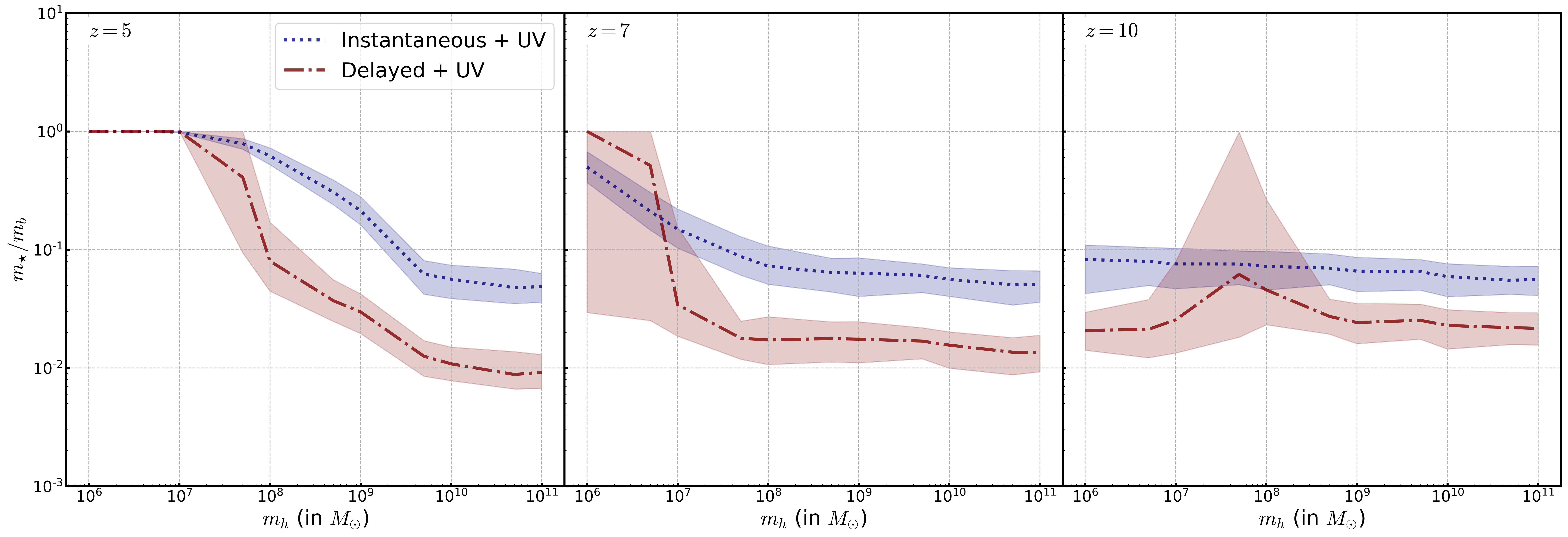}  
		\caption{Baryon mass assembly history of a $m_\text{h}=10^7\text{M}_{\odot}$ at $z$=5. The upper and lower panels show $m_\text{g}$ and $m_\star$ (in $\text{M}_{\odot}$) against cosmic time (in Gyrs); solid (dotted) curves correspond to instantaneous star formation and feedback without (with) UV suppression of accretion, while dashed (dot-dashed) curves correspond to bursty star formation without (with) UV suppression of accretion.}
		\label{fig:StarToBaryonMassFraction}
	\end{figure*}

 In Figure~\ref{fig:StarToBaryonMassFraction} we quantify how $m_\text{g}$ and $m_\star$ vary with $m_\text{h}$, looking at three epochs - $z$=10, 7, and 5 (from right to left). Because we are interested in relative trends, we plot the quantity $m_\star/(m_\star+m_\text{g})$ against $m_\text{h}$ at that epoch. As the system becomes gas depleted, we expect that $m_\star/(m_\star+m_\text{g})\rightarrow 1$ because $m_\text{g}\rightarrow 0$, which we expect at lower halo masses and should move to a higher mass scale following the onset of the UV suppression of gas accretion from the IGM. 
        
    \par At $z$=10, we see that this ratio shows little variation across the mass range, with the median $m_\star/(m_\star+m_\text{g})\simeq 0.05$ in the instantaneous + UV case, compared to $m_\star/(m_\star+m_\text{g})\simeq 0.03$ in the delayed + UV case. There is a weak trend for $m_\star/(m_\star+m_\text{g})$ to decrease (increase) with increasing mass in the instantaneous (delayed) + UV case. Interestingly, we see an enhancement $m_\star/(m_\star+m_\text{g})$ in the delayed + UV case in the halo mass range $10^7 \text{M}_\odot \leq m_\text{h} \leq 10^8 \text{M}_\odot$ - this reflects the transitional nature of this halo mass range and how halo mass assembly histories interact with burstiness in star formation and delayed feedback to produce a range in gas mass.

    \par By $z$=7, we now see a marked variation across the mass range. At higher masses, the median $m_\star/(m_\star+m_\text{g})\simeq 0.05$ in the instantaneous + UV case, unchanged from $z$=10; in contrast, $m_\star/(m_\star+m_\text{g})\simeq 0.02$ in the delayed + UV case, which has is lower than at $z$=10. The ratio increases smoothly towards lower masses in the instantaneous + UV case, whereas the ratio is flat down to $m_\text{h} \simeq 10^7 \text{M}_\odot$ before sharply turning up to $m_\star/(m_\star+m_\text{g})\simeq 1$. At $z$=5, the marked variation with mass has strengthened. At higher masses, the median $m_\star/(m_\star+m_\text{g})\simeq 0.05$ in the instantaneous + UV case is unchanged from $z$=10, whereas $m_\star/(m_\star+m_\text{g})\simeq 0.01$ in the delayed + UV case, which is marginally lower than at $z$=7. We now see a gradual rise in $m_\star/(m_\star+m_\text{g})$ towards lower $m_\text{h}$ becoming evident at $m_\text{h} \simeq 10^{10} \text{M}_\odot$; by $m_\star/(m_\star+m_\text{g})$=1 by $m_\text{h} \simeq 10^7 \text{M}_\odot$ in both the instantaneous + UV and delayed + UV cases.
    
    \par At the three epochs we have considered, we see a general trend for $m_\star/(m_\star+m_\text{g})$ to be approximately a factor of $3-5$ larger in the instantaneous + UV case compared to the delayed + UV case. Depending on epoch and mass scale, this can mean both a relative enhancement in stellar mass or a relative deficit in gas mass in the instantaneous + UV model. However, the trend is for the lower mass galaxies to have higher stellar masses in the delayed + UV case, while - in the context of this particular physical model - higher mass galaxies can retain their gas reservoirs. Physically, we would expect seed super-massive black hole growth and associated feedback to reduce gas mass in these high mass galaxies. We will revisit this in a future paper.

	\section{Conclusions}
    \label{sec:conclusions}
    We have investigated how the baryon mass content (i.e. the stellar and gas mass) of dark matter halos in the early Universe ($z \geq 5$) is affected by the complex interplay between bursty star formation - arising from delayed feedback - and the UV suppression of gas accretion from the IGM driven by cosmological reionization. Using a lightweight semi-analytical model that we have written, we have assessed how global factors - the emergence of an ionising ultraviolet (UV) radiation background - and local factors - star formation and feedback efficiency, time delayed feedback, and variations in dark matter halo assembly histories - influence the evolution of baryon mass content as a function of cosmic time ($z \geq 5$) and dark matter halo mass ($10^6 \text{M}_\odot \leq m_\text{h} \leq 10^{11} \text{M}_\odot$).

    \par Using the assembly histories of baryons in a sample of halos with $m_\text{h}=10^7 \text{M}_\odot$ at $z$=5 to illustrate the relative importance of physical processes, we investigated how star formation efficiency, the strength of feedback, and the delayed feedback timescale influenced the stellar and gas masses over cosmic time. As shown by previous studies, delayed feedback leads to oscillations in gas mass, evident more subtly in the stellar mass, as a function of cosmic time. If star formation is inefficient, gas and stellar mass growth histories are similar regardless of whether or not feedback is instantaneous or delayed; in contrast, highly efficient star formation can drive strong oscillations, as we would anticipate given that this population produces feedback via supernovae winds. Weaker delayed feedback cannot suppress the oscillatory behaviour in lower mass systems and these systems can retain their gas for a more extended period, whereas the main effect of stronger feedback is to drive $m_\text{g} \rightarrow 0$ more quickly. Longer delayed feedback timescales reduce the number of oscillations in $m_\text{g}$ and can result in more gas being retained at later times, whereas an earlier redshift of reionization has the simple effect of nudging the evolutionary trends in $m_\text{g}$ and $m_\star$ (i.e. $m_\text{g} \rightarrow 0)$, plateauing of $m_\star$ to the correspondingly earlier time.

    \par If we consider the variation of baryon mass content - which we parameterise by the ratio $m_\star/(m_\star+m_\text{g})$ - with halo mass, we find at earlier times that the median trend is for the ratio to be relatively flat and small in both the instantaneous and delayed feedback cases - of order $0.01-0.1$ - but with $m_\star/(m_\star+m_\text{g})$ to be approximately a factor of $3-5$ larger in the instantaneous case. At later times, we find that the ratio increases with decreasing halo mass such that $m_\star/(m_\star+m_\text{g})\rightarrow 1$ in the lowest mass systems. Higher mass systems retain their gas reservoirs in both cases considered because they are sufficiently massive to continue accreting gas from the IGM in a way that less massive systems cannot during cosmological reionization. 
    
    \par The relative gas richness of massive systems is at odds with both observations and theoretical expectations \citep[e.g.][]{Moster2018,Labbe2023}, but we would expect feedback from super-massive black holes in these massive galaxies to expel gas and quench star formation. As noted in \S~\ref{sec:model}, we do not model seed supermassive black growth or its associated feedback; we expect this to be important \citep[e.g.][]{Power2011} in an interesting, mass scale dependent manner \citep[cf.][]{Nayakshin2009MNRAS} and we shall explore it in a future paper.

    \par We have not explicitly considered the influence of metallicity in this paper, although we have investigated straightforward extensions to our model to do this \citep[e.g.][]{Kravtsov2022}. We expect that feedback efficiency should depend on metallicity \citep[e.g.][]{Jecmen2023,Sugimura2024}. For example, the ability of massive stars to drive stellar winds depends on the interaction cross-section of energetic photons with gas in stellar atmospheres, which depends on the presence of heavier elements \citep[cf.][]{Lamers1999}. This will introduce another mass-scale dependence, reflecting the number of generations of stars and metal enrichment events, and we shall explore it in future work.

    \section{Acknowledgement}
    AM and CP thank the anonymous referee for their very positive report.
    CP acknowledges the support of the ARC Centre of Excellence for All Sky Astrophysics in 3 Dimensions (ASTRO 3D), through project number CE170100013.

    \appendix

    \section{Evolution of Baryon Mass Content at Low and High Halo Masses}
    \label{sec:appendix_massdependence}
\begin{figure}[!h]
		\centering
		\includegraphics[width=\linewidth]{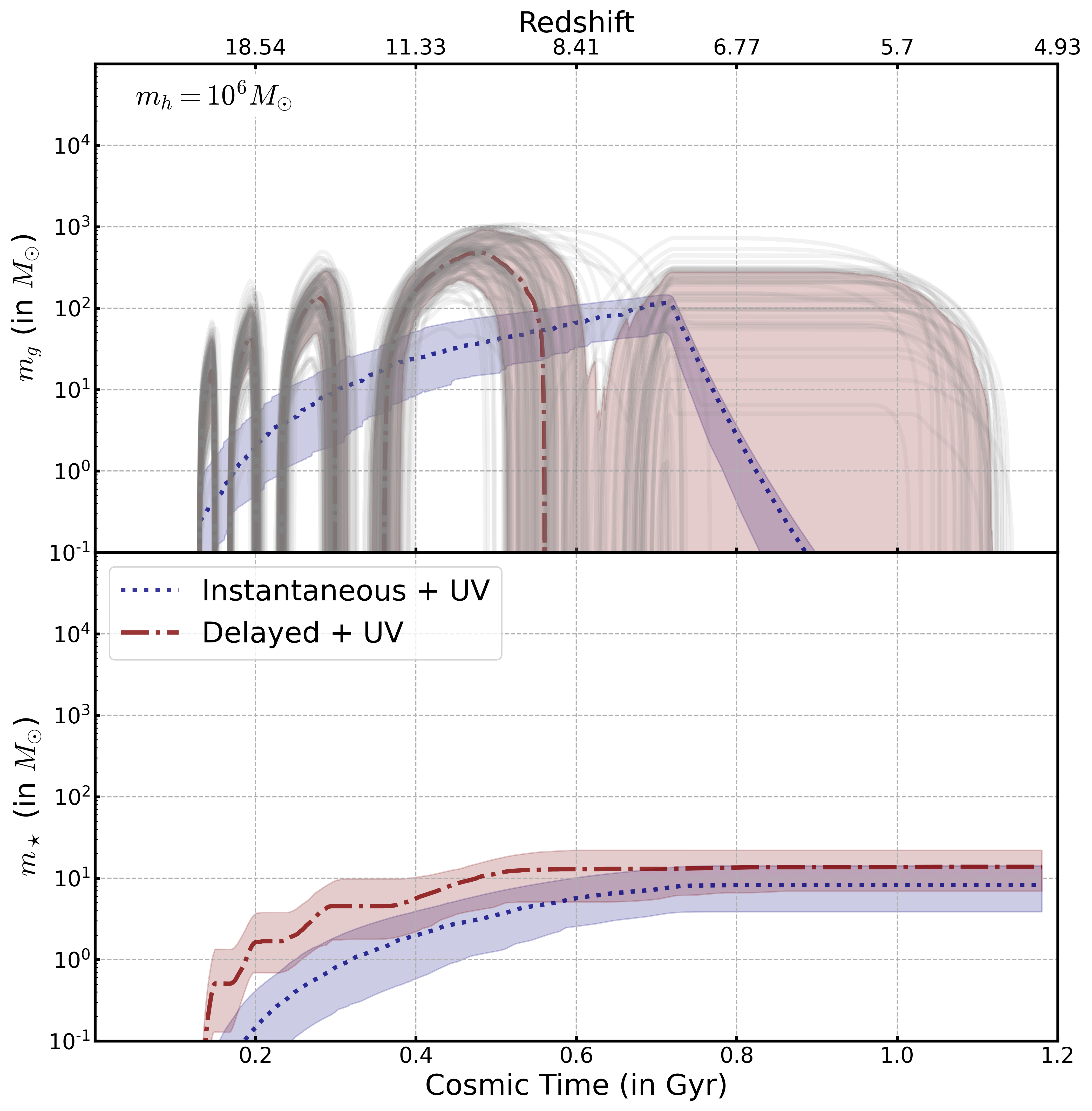}
  		\includegraphics[width=\linewidth]{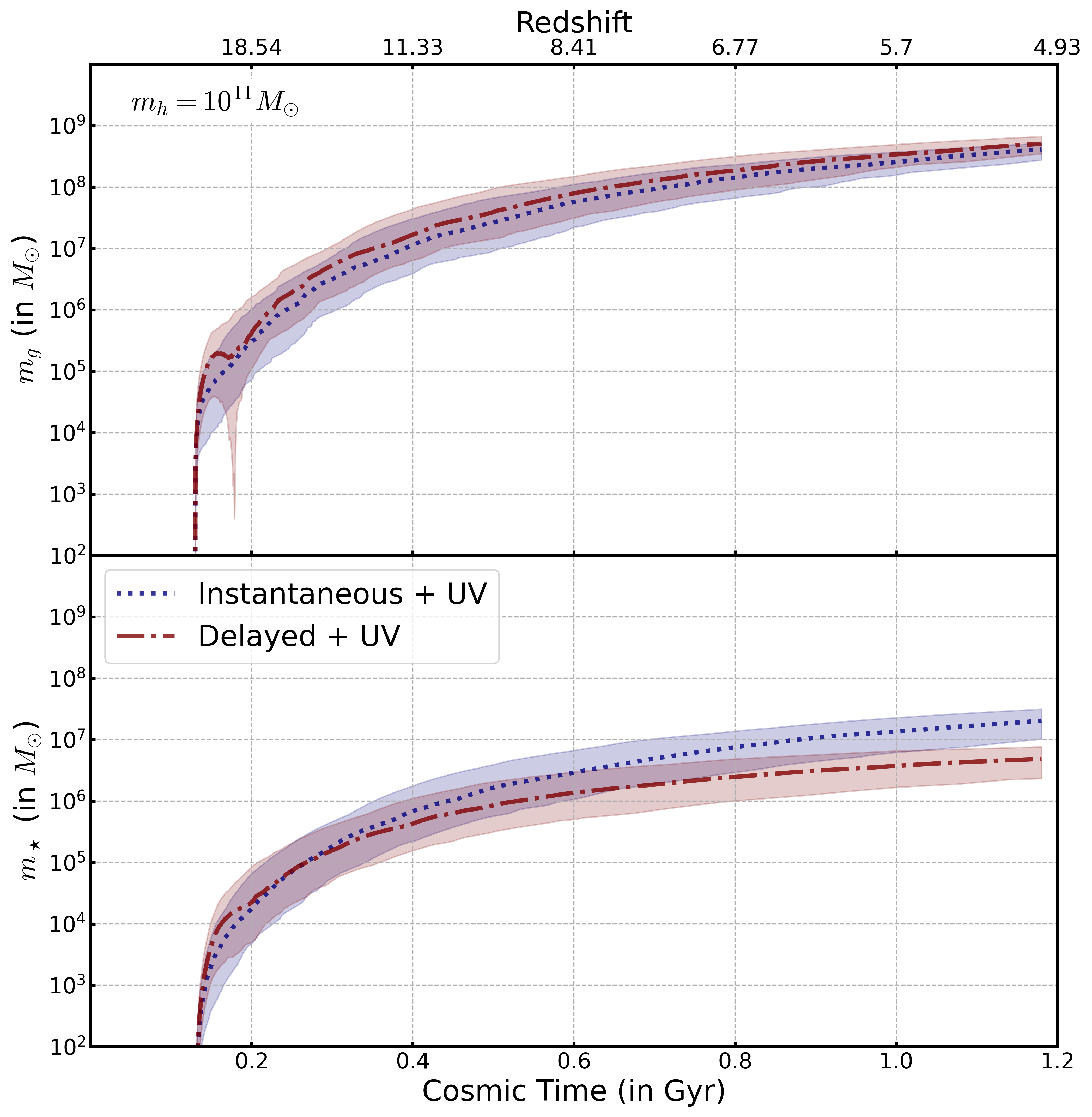}
		\caption{Baryon mass assembly history of a sample of 100 halos with $m_\text{h}=10^6\text{M}_{\odot}$ (upper panel) and $m_\text{h}=10^{11}\text{M}_{\odot}$ (lower panel) at $z$=5. The upper and lower panels show the median values of $m_\text{g}$ and $m_\star$ (in $\text{M}_{\odot}$) against cosmic time (in Gyrs); dotted curves correspond to instantaneous star formation and feedback with UV suppression of accretion, while dot-dashed curves correspond to bursty star formation with UV suppression of accretion. The coloured bands indicate the range of the 10$^\text{th}$ and 90$^\text{th}$ percentiles.}
		\label{fig:BaryonAssemblyHistoryPopulation_Appendix}
	\end{figure}

   In Figure~\ref{fig:BaryonAssemblyHistoryPopulation_Appendix}, we show the variation in $m_\text{g}$ and $m_\star$ with cosmic time for a sample of 100 halos, each with $m_\text{h}=10^7\text{M}_{\odot}$ at $z=5$. The dotted (dot-dashed) curves correspond to the median values of $m_\text{g}$ (upper panel) and $m_\star$ (lower) for instantaneous (bursty) star formation and feedback with UV suppression of accretion, while the coloured bands indicate the range of the 10$^\text{th}$ and 90$^\text{th}$ percentiles. This shows how variations in the assembly history of the underlying dark matter halo, whose growth rate $\dot{m}_\text{h}$ governs the growth rate of the gas mass and consequently the stellar mass. Interestingly we see a large variation in gas mass at late times for the lower mass system - although the median $m_\text{g}=0$, there is a large number of systems with $m_\text{g}=10-100 \text{M}_\odot$.

    \printbibliography

@Article{Murray2011,
	author    = {Norman Murray},
	journal   = {The Astrophysical Journal},
	title     = {Star formation efficiencies and lifetimes of giant molecular clouds in the Milky Way},
	year      = {2011},
	month     = {2},
	number    = {2},
	pages     = {133},
	volume    = {729},
	doi       = {10.1088/0004-637x/729/2/133},
	publisher = {American Astronomical Society},
}

@Article{Kravtsov2022,
	author    = {Andrey Kravtsov and Viraj Manwadkar},
	journal   = {Monthly Notices of the Royal Astronomical Society},
	title     = {GRUMPY: a simple framework for realistic forward modelling of dwarf galaxies},
	year      = {2022},
	month     = {5},
	number    = {2},
	pages     = {2667--2691},
	volume    = {514},
	doi       = {10.1093/mnras/stac1439},
	publisher = {Oxford University Press ({OUP})},
}

@Article{Furlanetto2022,
	author    = {Steven R. Furlanetto and Jordan Mirocha},
	journal   = {Monthly Notices of the Royal Astronomical Society},
	title     = {Bursty star formation during the Cosmic Dawn driven by delayed stellar feedback},
	year      = {2022},
	month     = {2},
	number    = {3},
	pages     = {3895--3909},
	volume    = {511},
	doi       = {10.1093/mnras/stac310},
	publisher = {Oxford University Press ({OUP})},
}

@Article{Efstathiou1992,
	author    = {G. Efstathiou},
	journal   = {Monthly Notices of the Royal Astronomical Society},
	title     = {Suppressing the formation of dwarf galaxies via photoionization},
	year      = {1992},
	month     = {5},
	number    = {1},
	pages     = {43P--47P},
	volume    = {256},
	doi       = {10.1093/mnras/256.1.43p},
	publisher = {Oxford University Press ({OUP})},
}

@Article{Gnedin2000,
	author    = {Nickolay Y. Gnedin},
	journal   = {The Astrophysical Journal},
	title     = {Effect of Reionization on Structure Formation in the Universe},
	year      = {2000},
	month     = {10},
	number    = {2},
	pages     = {535--541},
	volume    = {542},
	doi       = {10.1086/317042},
	publisher = {American Astronomical Society},
}

@Article{Okamoto2008,
	author    = {Takashi Okamoto and Liang Gao and Tom Theuns},
	journal   = {Monthly Notices of the Royal Astronomical Society},
	title     = {Mass loss of galaxies due to an ultraviolet background},
	year      = {2008},
	month     = {11},
	number    = {3},
	pages     = {920--928},
	volume    = {390},
	doi       = {10.1111/j.1365-2966.2008.13830.x},
	publisher = {Oxford University Press ({OUP})},
}

@Article{Furlanetto2017,
	author    = "Steven R. Furlanetto and Jordan Mirocha and Richard H. Mebane and Guochao Sun",
	journal   = {Monthly Notices of the Royal Astronomical Society},
	title     = {A minimalist feedback-regulated model for galaxy formation during the epoch of reionization},
	year      = {2017},
	month     = {8},
	number    = {2},
	pages     = {1576--1592},
	volume    = {472},
	doi       = {10.1093/mnras/stx2132},
	publisher = {Oxford University Press ({OUP})},
}

@Article{FaucherGiguere2018,
	author    = {Claude-Andr{\'{e}} Faucher-Gigu{\`{e}}re},
	journal   = {Monthly Notices of the Royal Astronomical Society},
	title     = {A model for the origin of bursty star formation in galaxies},
	year      = {2018},
	month     = {1},
	number    = {3},
	pages     = {3717--3731},
	volume    = {473},
	doi       = {10.1093/mnras/stx2595},
	publisher = {Oxford University Press ({OUP})},
}

@Article{Leroy2017,
	author    = {Adam K. Leroy and Eva Schinnerer and Annie Hughes and J. M. Diederik Kruijssen and Sharon Meidt and Andreas Schruba and Jiayi Sun and Frank Bigiel and Gonzalo Aniano and Guillermo A. Blanc and Alberto Bolatto and M{\'{e}}lanie Chevance and Dario Colombo and Molly Gallagher and Santiago Garcia-Burillo and Carsten Kramer and Miguel Querejeta and Jerome Pety and Todd A. Thompson and Antonio Usero},
	journal   = {The Astrophysical Journal},
	title     = {Cloud-scale {ISM} Structure and Star Formation in M51},
	year      = {2017},
	month     = {8},
	number    = {1},
	pages     = {71},
	volume    = {846},
	doi       = {10.3847/1538-4357/aa7fef},
	publisher = {American Astronomical Society},
}

@Article{Orr2019,
  author    = {Matthew E Orr and Christopher C Hayward and Philip F Hopkins},
  journal   = {Monthly Notices of the Royal Astronomical Society},
  title     = {A simple non-equilibrium feedback model for galaxy-scale star formation: delayed feedback and {SFR} scatter},
  year      = {2019},
  month     = {4},
  number    = {4},
  pages     = {4724--4737},
  volume    = {486},
  doi       = {10.1093/mnras/stz1156},
  publisher = {Oxford University Press ({OUP})},
}

@Article{Efstathiou2000,
  author    = {G. Efstathiou},
  journal   = {Monthly Notices of the Royal Astronomical Society},
  title     = {A model of supernova feedback in galaxy formation},
  year      = {2000},
  month     = {9},
  number    = {3},
  pages     = {697--719},
  volume    = {317},
  doi       = {10.1046/j.1365-8711.2000.03665.x},
  publisher = {Oxford University Press ({OUP})},
}

@Article{Scalo1986,
  author    = {J. M. Scalo and C. Struck-Marcell},
  journal   = {The Astrophysical Journal},
  title     = {A physical mechanism for bursts of star formation},
  year      = {1986},
  month     = {2},
  pages     = {77},
  volume    = {301},
  doi       = {10.1086/163874},
  publisher = {American Astronomical Society},
}

@Article{Faisst2019,
  author    = {Andreas L. Faisst and Peter L. Capak and Najmeh Emami and Sandro Tacchella and Kirsten L. Larson},
  journal   = {The Astrophysical Journal},
  title     = {The Recent Burstiness of Star Formation in Galaxies at $<$i$>$z$<$/i$>$ $\sim$ 4.5 from H$<$i$>$$\alpha$$<$/i$>$ Measurements},
  year      = {2019},
  month     = {10},
  number    = {2},
  pages     = {133},
  volume    = {884},
  doi       = {10.3847/1538-4357/ab425b},
  publisher = {American Astronomical Society},
}

@Article{Strait2023,
  author    = {Victoria Strait and Gabriel Brammer and Adam Muzzin and Guillaume Desprez and Yoshihisa Asada and Roberto Abraham and Maru{\v{s}}a Brada{\v{c}} and Kartheik G. Iyer and Nicholas Martis and Lamiya Mowla and Gaël Noirot and Ghassan T. E. Sarrouh and Marcin Sawicki and Chris Willott and Katriona Gould and Tess Grindlay and Jasleen Matharu and Gregor Rihtar{\v{s}}i{\v{c}}},
  journal   = {The Astrophysical Journal Letters},
  title     = {An Extremely Compact, Low-mass Galaxy on its Way to Quiescence at z = 5.2},
  year      = {2023},
  month     = {5},
  number    = {2},
  pages     = {L23},
  volume    = {949},
  doi       = {10.3847/2041-8213/acd457},
  publisher = {American Astronomical Society},
}

@Misc{Looser2023,
  author    = {Looser, Tobias J. and D'Eugenio, Francesco and Maiolino, Roberto and Witstok, Joris and Sandles, Lester and Curtis-Lake, Emma and Chevallard, Jacopo and Tacchella, Sandro and Johnson, Benjamin D. and Baker, William M. and Suess, Katherine A. and Carniani, Stefano and Ferruit, Pierre and Arribas, Santiago and Bonaventura, Nina and Bunker, Andrew J. and Cameron, Alex J. and Charlot, Stephane and Curti, Mirko and de Graaff, Anna and Maseda, Michael V. and Rawle, Tim and Rix, Hans-Walter and Del Pino, Bruno Rodriguez and Smit, Renske and Übler, Hannah and Willott, Chris and Alberts, Stacey and Egami, Eiichi and Eisenstein, Daniel J. and Endsley, Ryan and Hausen, Ryan and Rieke, Marcia and Robertson, Brant and Shivaei, Irene and Williams, Christina C. and Boyett, Kristan and Chen, Zuyi and Ji, Zhiyuan and Jones, Gareth J. and Kumari, Nimisha and Nelson, Erica and Perna, Michele and Saxena, Aayush and Scholtz, 1},
  title     = {Discovery of a quiescent galaxy at z=7.3},
  year      = {2023},
  copyright = {Creative Commons Attribution 4.0 International},
  doi       = {10.48550/ARXIV.2302.14155},
  publisher = {arXiv},
}

@Misc{Dome2023,
  author    = {Dome, Tibor and Tacchella, Sandro and Fialkov, Anastasia and Dekel, Avishai and Ginzburg, Omri and Lapiner, Sharon and Looser, Tobias J.},
  title     = {Mini-Quenching of High-Redshift Galaxies by Bursty Star Formation},
  year      = {2023},
  copyright = {Creative Commons Attribution 4.0 International},
  doi       = {10.48550/ARXIV.2305.07066},
  publisher = {arXiv},
}

@Article{Emami2019,
  author    = {Najmeh Emami and Brian Siana and Daniel R. Weisz and Benjamin D. Johnson and Xiangcheng Ma and Kareem El-Badry},
  journal   = {The Astrophysical Journal},
  title     = {A Closer Look at Bursty Star Formation with $L_{\text{H}\alpha}$ and $L_\text{UV}$ Distributions},
  year      = {2019},
  month     = {8},
  number    = {1},
  pages     = {71},
  volume    = {881},
  doi       = {10.3847/1538-4357/ab211a},
  publisher = {American Astronomical Society},
}

@Article{Onorbe2015,
  author    = {Jose O{\~{n}}orbe and Michael Boylan-Kolchin and James S. Bullock and Philip F. Hopkins and Du{\v{s}}an Kere{\v{s}} and Claude-Andr{\'{e}} Faucher-Gigu{\`{e}}re and Eliot Quataert and Norman Murray},
  journal   = {Monthly Notices of the Royal Astronomical Society},
  title     = {Forged in fire: cusps, cores and baryons in low-mass dwarf galaxies},
  year      = {2015},
  month     = {10},
  number    = {2},
  pages     = {2092--2106},
  volume    = {454},
  doi       = {10.1093/mnras/stv2072},
  publisher = {Oxford University Press ({OUP})},
}

@ARTICLE{boylan-kolchin2023,
       author = {{Boylan-Kolchin}, Michael},
        title = "{Stress testing {\ensuremath{\Lambda}}CDM with high-redshift galaxy candidates}",
      journal = {Nature Astronomy},
         year = 2023,
        month = jun,
       volume = {7},
        pages = {731-735},
          doi = {10.1038/s41550-023-01937-7},
archivePrefix = {arXiv},
       eprint = {2208.01611},
 primaryClass = {astro-ph.CO},
       adsurl = {https://ui.adsabs.harvard.edu/abs/2023NatAs...7..731B}
}

@ARTICLE{KraghJespersen2024,
       author = {{Kragh Jespersen}, Christian and {Steinhardt}, Charles L. and {Somerville}, Rachel S. and {Lovell}, Christopher C.},
        title = "{On the Significance of Rare Objects at High Redshift: The Impact of Cosmic Variance}",
      journal = {arXiv e-prints},
         year = 2024,
        month = feb,
          eid = {arXiv:2403.00050},
        pages = {arXiv:2403.00050},
          doi = {10.48550/arXiv.2403.00050},
archivePrefix = {arXiv},
       eprint = {2403.00050},
 primaryClass = {astro-ph.GA},
       adsurl = {https://ui.adsabs.harvard.edu/abs/2024arXiv240300050K}
}

@ARTICLE{Finkelstein2023,
       author = {{Finkelstein}, Steven L. and {Bagley}, Micaela B. and {Ferguson}, Henry C. and {Wilkins}, Stephen M. and {Kartaltepe}, Jeyhan S. and {Papovich}, Casey and {Yung}, L.~Y. Aaron and {Arrabal Haro}, Pablo and {Behroozi}, Peter and {Dickinson}, Mark and {Kocevski}, Dale D. and {Koekemoer}, Anton M. and {Larson}, Rebecca L. and {Le Bail}, Aur{\'e}lien and {Morales}, Alexa M. and {P{\'e}rez-Gonz{\'a}lez}, Pablo G. and {Burgarella}, Denis and {Dav{\'e}}, Romeel and {Hirschmann}, Michaela and {Somerville}, Rachel S. and {Wuyts}, Stijn and {Bromm}, Volker and {Casey}, Caitlin M. and {Fontana}, Adriano and {Fujimoto}, Seiji and {Gardner}, Jonathan P. and {Giavalisco}, Mauro and {Grazian}, Andrea and {Grogin}, Norman A. and {Hathi}, Nimish P. and {Hutchison}, Taylor A. and {Jha}, Saurabh W. and {Jogee}, Shardha and {Kewley}, Lisa J. and {Kirkpatrick}, Allison and {Long}, Arianna S. and {Lotz}, Jennifer M. and {Pentericci}, Laura and {Pierel}, Justin D.~R. and {Pirzkal}, Nor and {Ravindranath}, Swara and {Ryan}, Russell E. and {Trump}, Jonathan R. and {Yang}, Guang and {Bhatawdekar}, Rachana and {Bisigello}, Laura and {Buat}, V{\'e}ronique and {Calabr{\`o}}, Antonello and {Castellano}, Marco and {Cleri}, Nikko J. and {Cooper}, M.~C. and {Croton}, Darren and {Daddi}, Emanuele and {Dekel}, Avishai and {Elbaz}, David and {Franco}, Maximilien and {Gawiser}, Eric and {Holwerda}, Benne W. and {Huertas-Company}, Marc and {Jaskot}, Anne E. and {Leung}, Gene C.~K. and {Lucas}, Ray A. and {Mobasher}, Bahram and {Pandya}, Viraj and {Tacchella}, Sandro and {Weiner}, Benjamin J. and {Zavala}, Jorge A.},
        title = "{CEERS Key Paper. I. An Early Look into the First 500 Myr of Galaxy Formation with JWST}",
      journal = {\apjl},
         year = 2023,
        month = mar,
       volume = {946},
       number = {1},
          eid = {L13},
        pages = {L13},
          doi = {10.3847/2041-8213/acade4},
archivePrefix = {arXiv},
       eprint = {2211.05792},
 primaryClass = {astro-ph.GA},
       adsurl = {https://ui.adsabs.harvard.edu/abs/2023ApJ...946L..13F}
}

@ARTICLE{Finkelstein2023_2,
       author = {{Finkelstein}, Steven L. and {Leung}, Gene C.~K. and {Bagley}, Micaela B. and {Dickinson}, Mark and {Ferguson}, Henry C. and {Papovich}, Casey and {Akins}, Hollis B. and {Arrabal Haro}, Pablo and {Dave}, Romeel and {Dekel}, Avishai and {Kartaltepe}, Jeyhan S. and {Kocevski}, Dale D. and {Koekemoer}, Anton M. and {Pirzkal}, Norbert and {Somerville}, Rachel S. and {Yung}, L.~Y. Aaron and {Amorin}, Ricardo and {Backhaus}, Bren E. and {Behroozi}, Peter and {Bisigello}, Laura and {Bromm}, Volker and {Casey}, Caitlin M. and {Chavez Ortiz}, Oscar A. and {Cheng}, Yingjie and {Chworowsky}, Katherine and {Cleri}, Nikko J. and {Cooper}, Michael C. and {Davis}, Kelcey and {de la Vega}, Alexander and {Elbaz}, David and {Franco}, Maximilien and {Fontana}, Adriano and {Fujimoto}, Seiji and {Giavalisco}, Mauro and {Grogin}, Norman A. and {Holwerda}, Benne W. and {Huertas-Company}, Marc and {Hirschmann}, Michaela and {Iyer}, Kartheik G. and {Jogee}, Shardha and {Jung}, Intae and {Larson}, Rebecca L. and {Lucas}, Ray A. and {Mobasher}, Bahram and {Morales}, Alexa M. and {Morley}, Caroline V. and {Mukherjee}, Sagnick and {Perez-Gonzalez}, Pablo G. and {Ravindranath}, Swara and {Rodighiero}, Giulia and {Rowland}, Melanie and {Tacchella}, Sandro and {Taylor}, Anthony J. and {Trump}, Jonathan R. and {Wilkins}, Stephen},
        title = "{The Complete CEERS Early Universe Galaxy Sample: A Surprisingly Slow Evolution of the Space Density of Bright Galaxies at z \raisebox{-0.5ex}\textasciitilde 8.5-14.5}",
      journal = {arXiv e-prints},
         year = 2023,
        month = nov,
          eid = {arXiv:2311.04279},
        pages = {arXiv:2311.04279},
          doi = {10.48550/arXiv.2311.04279},
archivePrefix = {arXiv},
       eprint = {2311.04279},
 primaryClass = {astro-ph.GA},
       adsurl = {https://ui.adsabs.harvard.edu/abs/2023arXiv231104279F}
}

@ARTICLE{Dekel2023,
       author = {{Dekel}, Avishai and {Sarkar}, Kartick C. and {Birnboim}, Yuval and {Mandelker}, Nir and {Li}, Zhaozhou},
        title = "{Efficient formation of massive galaxies at cosmic dawn by feedback-free starbursts}",
      journal = {\mnras},
         year = 2023,
        month = aug,
       volume = {523},
       number = {3},
        pages = {3201-3218},
          doi = {10.1093/mnras/stad1557},
archivePrefix = {arXiv},
       eprint = {2303.04827},
 primaryClass = {astro-ph.GA},
       adsurl = {https://ui.adsabs.harvard.edu/abs/2023MNRAS.523.3201D}
}

@ARTICLE{Dave2012,
       author = {{Dav{\'e}}, Romeel and {Finlator}, Kristian and {Oppenheimer}, Benjamin D.},
        title = "{An analytic model for the evolution of the stellar, gas and metal content of galaxies}",
      journal = {\mnras},
         year = 2012,
        month = mar,
       volume = {421},
       number = {1},
        pages = {98-107},
          doi = {10.1111/j.1365-2966.2011.20148.x},
archivePrefix = {arXiv},
       eprint = {1108.0426},
 primaryClass = {astro-ph.CO},
       adsurl = {https://ui.adsabs.harvard.edu/abs/2012MNRAS.421...98D}
}

@ARTICLE{Labbe2023,
       author = {{Labb{\'e}}, Ivo and {van Dokkum}, Pieter and {Nelson}, Erica and {Bezanson}, Rachel and {Suess}, Katherine A. and {Leja}, Joel and {Brammer}, Gabriel and {Whitaker}, Katherine and {Mathews}, Elijah and {Stefanon}, Mauro and {Wang}, Bingjie},
        title = "{A population of red candidate massive galaxies  600 Myr after the Big Bang}",
      journal = {\nat},
         year = 2023,
        month = apr,
       volume = {616},
       number = {7956},
        pages = {266-269},
          doi = {10.1038/s41586-023-05786-2},
archivePrefix = {arXiv},
       eprint = {2207.12446},
 primaryClass = {astro-ph.GA},
       adsurl = {https://ui.adsabs.harvard.edu/abs/2023Natur.616..266L}
}

@ARTICLE{thoul1996,
       author = {{Thoul}, Anne A. and {Weinberg}, David H.},
        title = "{Hydrodynamic Simulations of Galaxy Formation. II. Photoionization and the Formation of Low-Mass Galaxies}",
      journal = {\apj},
         year = 1996,
        month = jul,
       volume = {465},
        pages = {608},
          doi = {10.1086/177446},
archivePrefix = {arXiv},
       eprint = {astro-ph/9510154},
 primaryClass = {astro-ph},
       adsurl = {https://ui.adsabs.harvard.edu/abs/1996ApJ...465..608T}
}

@ARTICLE{Dekel1986,
       author = {{Dekel}, A. and {Silk}, J.},
        title = "{The Origin of Dwarf Galaxies, Cold Dark Matter, and Biased Galaxy Formation}",
      journal = {\apj},
         year = 1986,
        month = apr,
       volume = {303},
        pages = {39},
          doi = {10.1086/164050},
       adsurl = {https://ui.adsabs.harvard.edu/abs/1986ApJ...303...39D}
}

@ARTICLE{Sparre2017,
       author = {{Sparre}, Martin and {Hayward}, Christopher C. and {Feldmann}, Robert and {Faucher-Gigu{\`e}re}, Claude-Andr{\'e} and {Muratov}, Alexander L. and {Kere{\v{s}}}, Du{\v{s}}an and {Hopkins}, Philip F.},
        title = "{(Star)bursts of FIRE: observational signatures of bursty star formation in galaxies}",
      journal = {\mnras},
         year = 2017,
        month = apr,
       volume = {466},
       number = {1},
        pages = {88-104},
          doi = {10.1093/mnras/stw3011},
archivePrefix = {arXiv},
       eprint = {1510.03869},
 primaryClass = {astro-ph.GA},
       adsurl = {https://ui.adsabs.harvard.edu/abs/2017MNRAS.466...88S}
}

@ARTICLE{Muratov2015,
       author = {{Muratov}, Alexander L. and {Kere{\v{s}}}, Du{\v{s}}an and {Faucher-Gigu{\`e}re}, Claude-Andr{\'e} and {Hopkins}, Philip F. and {Quataert}, Eliot and {Murray}, Norman},
        title = "{Gusty, gaseous flows of FIRE: galactic winds in cosmological simulations with explicit stellar feedback}",
      journal = {\mnras},
         year = 2015,
        month = dec,
       volume = {454},
       number = {3},
        pages = {2691-2713},
          doi = {10.1093/mnras/stv2126},
archivePrefix = {arXiv},
       eprint = {1501.03155},
 primaryClass = {astro-ph.GA},
       adsurl = {https://ui.adsabs.harvard.edu/abs/2015MNRAS.454.2691M}
}

@ARTICLE{Wyithe2006,
       author = {{Wyithe}, J. Stuart B. and {Loeb}, Abraham},
        title = "{Suppression of dwarf galaxy formation by cosmic reionization}",
      journal = {\nat},
         year = 2006,
        month = may,
       volume = {441},
       number = {7091},
        pages = {322-324},
          doi = {10.1038/nature04748},
archivePrefix = {arXiv},
       eprint = {astro-ph/0603550},
 primaryClass = {astro-ph},
       adsurl = {https://ui.adsabs.harvard.edu/abs/2006Natur.441..322W}
}

@ARTICLE{Wyithe2003,
       author = {{Wyithe}, J. Stuart B. and {Loeb}, Abraham},
        title = "{Reionization of Hydrogen and Helium by Early Stars and Quasars}",
      journal = {\apj},
         year = 2003,
        month = apr,
       volume = {586},
       number = {2},
        pages = {693-708},
          doi = {10.1086/367721},
archivePrefix = {arXiv},
       eprint = {astro-ph/0209056},
 primaryClass = {astro-ph},
       adsurl = {https://ui.adsabs.harvard.edu/abs/2003ApJ...586..693W}
}

@ARTICLE{Barkana2001,
       author = {{Barkana}, R. and {Loeb}, A.},
        title = "{In the beginning: the first sources of light and the reionization of the universe}",
      journal = {\physrep},
         year = 2001,
        month = jul,
       volume = {349},
       number = {2},
        pages = {125-238},
          doi = {10.1016/S0370-1573(01)00019-9},
archivePrefix = {arXiv},
       eprint = {astro-ph/0010468},
 primaryClass = {astro-ph},
       adsurl = {https://ui.adsabs.harvard.edu/abs/2001PhR...349..125B}
}

@ARTICLE{Weisz2012,
       author = {{Weisz}, Daniel R. and {Johnson}, Benjamin D. and {Johnson}, L. Clifton and {Skillman}, Evan D. and {Lee}, Janice C. and {Kennicutt}, Robert C. and {Calzetti}, Daniela and {van Zee}, Liese and {Bothwell}, Matthew S. and {Dalcanton}, Julianne J. and {Dale}, Daniel A. and {Williams}, Benjamin F.},
        title = "{Modeling the Effects of Star Formation Histories on H{\ensuremath{\alpha}} and Ultraviolet Fluxes in nearby Dwarf Galaxies}",
      journal = {\apj},
         year = 2012,
        month = jan,
       volume = {744},
       number = {1},
          eid = {44},
        pages = {44},
          doi = {10.1088/0004-637X/744/1/44},
archivePrefix = {arXiv},
       eprint = {1109.2905},
 primaryClass = {astro-ph.CO},
       adsurl = {https://ui.adsabs.harvard.edu/abs/2012ApJ...744...44W}
}

@ARTICLE{Hopkins2023,
       author = {{Hopkins}, Philip F. and {Gurvich}, Alexander B. and {Shen}, Xuejian and {Hafen}, Zachary and {Grudi{\'c}}, Michael Y. and {Kurinchi-Vendhan}, Shalini and {Hayward}, Christopher C. and {Jiang}, Fangzhou and {Orr}, Matthew E. and {Wetzel}, Andrew and {Kere{\v{s}}}, Du{\v{s}}an and {Stern}, Jonathan and {Faucher-Gigu{\`e}re}, Claude-Andr{\'e} and {Bullock}, James and {Wheeler}, Coral and {El-Badry}, Kareem and {Loebman}, Sarah R. and {Moreno}, Jorge and {Boylan-Kolchin}, Michael and {Quataert}, Eliot},
        title = "{What causes the formation of discs and end of bursty star formation?}",
      journal = {\mnras},
         year = 2023,
        month = oct,
       volume = {525},
       number = {2},
        pages = {2241-2286},
          doi = {10.1093/mnras/stad1902},
archivePrefix = {arXiv},
       eprint = {2301.08263},
 primaryClass = {astro-ph.GA},
       adsurl = {https://ui.adsabs.harvard.edu/abs/2023MNRAS.525.2241H}
}

@ARTICLE{Parkinson2008,
       author = {{Parkinson}, Hannah and {Cole}, Shaun and {Helly}, John},
        title = "{Generating dark matter halo merger trees}",
      journal = {\mnras},
         year = 2008,
        month = jan,
       volume = {383},
       number = {2},
        pages = {557-564},
          doi = {10.1111/j.1365-2966.2007.12517.x},
archivePrefix = {arXiv},
       eprint = {0708.1382},
 primaryClass = {astro-ph},
       adsurl = {https://ui.adsabs.harvard.edu/abs/2008MNRAS.383..557P}
}

@ARTICLE{White1991,
       author = {{White}, Simon D.~M. and {Frenk}, Carlos S.},
        title = "{Galaxy Formation through Hierarchical Clustering}",
      journal = {\apj},
         year = 1991,
        month = sep,
       volume = {379},
        pages = {52},
          doi = {10.1086/170483},
       adsurl = {https://ui.adsabs.harvard.edu/abs/1991ApJ...379...52W}
}

@ARTICLE{Planck2018,
       author = {{Planck Collaboration} and {Aghanim}, N. and {Akrami}, Y. and {Ashdown}, M. and {Aumont}, J. and {Baccigalupi}, C. and {Ballardini}, M. and {Banday}, A.~J. and {Barreiro}, R.~B. and {Bartolo}, N. and {Basak}, S. and {Battye}, R. and {Benabed}, K. and {Bernard}, J. -P. and {Bersanelli}, M. and {Bielewicz}, P. and {Bock}, J.~J. and {Bond}, J.~R. and {Borrill}, J. and {Bouchet}, F.~R. and {Boulanger}, F. and {Bucher}, M. and {Burigana}, C. and {Butler}, R.~C. and {Calabrese}, E. and {Cardoso}, J. -F. and {Carron}, J. and {Challinor}, A. and {Chiang}, H.~C. and {Chluba}, J. and {Colombo}, L.~P.~L. and {Combet}, C. and {Contreras}, D. and {Crill}, B.~P. and {Cuttaia}, F. and {de Bernardis}, P. and {de Zotti}, G. and {Delabrouille}, J. and {Delouis}, J. -M. and {Di Valentino}, E. and {Diego}, J.~M. and {Dor{\'e}}, O. and {Douspis}, M. and {Ducout}, A. and {Dupac}, X. and {Dusini}, S. and {Efstathiou}, G. and {Elsner}, F. and {En{\ss}lin}, T.~A. and {Eriksen}, H.~K. and {Fantaye}, Y. and {Farhang}, M. and {Fergusson}, J. and {Fernandez-Cobos}, R. and {Finelli}, F. and {Forastieri}, F. and {Frailis}, M. and {Fraisse}, A.~A. and {Franceschi}, E. and {Frolov}, A. and {Galeotta}, S. and {Galli}, S. and {Ganga}, K. and {G{\'e}nova-Santos}, R.~T. and {Gerbino}, M. and {Ghosh}, T. and {Gonz{\'a}lez-Nuevo}, J. and {G{\'o}rski}, K.~M. and {Gratton}, S. and {Gruppuso}, A. and {Gudmundsson}, J.~E. and {Hamann}, J. and {Handley}, W. and {Hansen}, F.~K. and {Herranz}, D. and {Hildebrandt}, S.~R. and {Hivon}, E. and {Huang}, Z. and {Jaffe}, A.~H. and {Jones}, W.~C. and {Karakci}, A. and {Keih{\"a}nen}, E. and {Keskitalo}, R. and {Kiiveri}, K. and {Kim}, J. and {Kisner}, T.~S. and {Knox}, L. and {Krachmalnicoff}, N. and {Kunz}, M. and {Kurki-Suonio}, H. and {Lagache}, G. and {Lamarre}, J. -M. and {Lasenby}, A. and {Lattanzi}, M. and {Lawrence}, C.~R. and {Le Jeune}, M. and {Lemos}, P. and {Lesgourgues}, J. and {Levrier}, F. and {Lewis}, A. and {Liguori}, M. and {Lilje}, P.~B. and {Lilley}, M. and {Lindholm}, V. and {L{\'o}pez-Caniego}, M. and {Lubin}, P.~M. and {Ma}, Y. -Z. and {Mac{\'\i}as-P{\'e}rez}, J.~F. and {Maggio}, G. and {Maino}, D. and {Mandolesi}, N. and {Mangilli}, A. and {Marcos-Caballero}, A. and {Maris}, M. and {Martin}, P.~G. and {Martinelli}, M. and {Mart{\'\i}nez-Gonz{\'a}lez}, E. and {Matarrese}, S. and {Mauri}, N. and {McEwen}, J.~D. and {Meinhold}, P.~R. and {Melchiorri}, A. and {Mennella}, A. and {Migliaccio}, M. and {Millea}, M. and {Mitra}, S. and {Miville-Desch{\^e}nes}, M. -A. and {Molinari}, D. and {Montier}, L. and {Morgante}, G. and {Moss}, A. and {Natoli}, P. and {N{\o}rgaard-Nielsen}, H.~U. and {Pagano}, L. and {Paoletti}, D. and {Partridge}, B. and {Patanchon}, G. and {Peiris}, H.~V. and {Perrotta}, F. and {Pettorino}, V. and {Piacentini}, F. and {Polastri}, L. and {Polenta}, G. and {Puget}, J. -L. and {Rachen}, J.~P. and {Reinecke}, M. and {Remazeilles}, M. and {Renzi}, A. and {Rocha}, G. and {Rosset}, C. and {Roudier}, G. and {Rubi{\~n}o-Mart{\'\i}n}, J.~A. and {Ruiz-Granados}, B. and {Salvati}, L. and {Sandri}, M. and {Savelainen}, M. and {Scott}, D. and {Shellard}, E.~P.~S. and {Sirignano}, C. and {Sirri}, G. and {Spencer}, L.~D. and {Sunyaev}, R. and {Suur-Uski}, A. -S. and {Tauber}, J.~A. and {Tavagnacco}, D. and {Tenti}, M. and {Toffolatti}, L. and {Tomasi}, M. and {Trombetti}, T. and {Valenziano}, L. and {Valiviita}, J. and {Van Tent}, B. and {Vibert}, L. and {Vielva}, P. and {Villa}, F. and {Vittorio}, N. and {Wandelt}, B.~D. and {Wehus}, I.~K. and {White}, M. and {White}, S.~D.~M. and {Zacchei}, A. and {Zonca}, A.},
        title = "{Planck 2018 results. VI. Cosmological parameters}",
      journal = {\aap},
         year = 2020,
        month = sep,
       volume = {641},
          eid = {A6},
        pages = {A6},
          doi = {10.1051/0004-6361/201833910},
archivePrefix = {arXiv},
       eprint = {1807.06209},
 primaryClass = {astro-ph.CO},
       adsurl = {https://ui.adsabs.harvard.edu/abs/2020A&A...641A...6P}
}

@ARTICLE{Bond1991,
       author = {{Bond}, J.~R. and {Cole}, S. and {Efstathiou}, G. and {Kaiser}, N.},
        title = "{Excursion Set Mass Functions for Hierarchical Gaussian Fluctuations}",
      journal = {\apj},
         year = 1991,
        month = oct,
       volume = {379},
        pages = {440},
          doi = {10.1086/170520},
       adsurl = {https://ui.adsabs.harvard.edu/abs/1991ApJ...379..440B}
}

@ARTICLE{Lacey1993,
       author = {{Lacey}, Cedric and {Cole}, Shaun},
        title = "{Merger rates in hierarchical models of galaxy formation}",
      journal = {\mnras},
         year = 1993,
        month = jun,
       volume = {262},
       number = {3},
        pages = {627-649},
          doi = {10.1093/mnras/262.3.627},
       adsurl = {https://ui.adsabs.harvard.edu/abs/1993MNRAS.262..627L}
}

@ARTICLE{Springel2005,
       author = {{Springel}, Volker and {White}, Simon D.~M. and {Jenkins}, Adrian and {Frenk}, Carlos S. and {Yoshida}, Naoki and {Gao}, Liang and {Navarro}, Julio and {Thacker}, Robert and {Croton}, Darren and {Helly}, John and {Peacock}, John A. and {Cole}, Shaun and {Thomas}, Peter and {Couchman}, Hugh and {Evrard}, August and {Colberg}, J{\"o}rg and {Pearce}, Frazer},
        title = "{Simulations of the formation, evolution and clustering of galaxies and quasars}",
      journal = {\nat},
         year = 2005,
        month = jun,
       volume = {435},
       number = {7042},
        pages = {629-636},
          doi = {10.1038/nature03597},
archivePrefix = {arXiv},
       eprint = {astro-ph/0504097},
 primaryClass = {astro-ph},
       adsurl = {https://ui.adsabs.harvard.edu/abs/2005Natur.435..629S}
}

@ARTICLE{Boyett2024,
       author = {{Boyett}, Kristan and {Trenti}, Michele and {Leethochawalit}, Nicha and {Calabr{\'o}}, Antonello and {Metha}, Benjamin and {Roberts-Borsani}, Guido and {Dalmasso}, Nicol{\'o} and {Yang}, Lilan and {Santini}, Paola and {Treu}, Tommaso and {Jones}, Tucker and {Henry}, Alaina and {Mason}, Charlotte A. and {Morishita}, Takahiro and {Nanayakkara}, Themiya and {Roy}, Namrata and {Wang}, Xin and {Fontana}, Adriano and {Merlin}, Emiliano and {Castellano}, Marco and {Paris}, Diego and {Brada{\v{c}}}, Maru{\v{s}}a and {Malkan}, Matt and {Marchesini}, Danilo and {Mascia}, Sara and {Glazebrook}, Karl and {Pentericci}, Laura and {Vanzella}, Eros and {Vulcani}, Benedetta},
        title = "{A massive interacting galaxy 510 million years after the Big Bang}",
      journal = {Nature Astronomy},
         year = 2024,
        month = mar,
          doi = {10.1038/s41550-024-02218-7},
       adsurl = {https://ui.adsabs.harvard.edu/abs/2024NatAs.tmp...56B}
}

@ARTICLE{Nayakshin2009MNRAS,
       author = {{Nayakshin}, Sergei and {Wilkinson}, Mark I. and {King}, Andrew},
        title = "{Competitive feedback in galaxy formation}",
      journal = {\mnras},
         year = 2009,
        month = sep,
       volume = {398},
       number = {1},
        pages = {L54-L57},
          doi = {10.1111/j.1745-3933.2009.00709.x},
archivePrefix = {arXiv},
       eprint = {0907.1002},
 primaryClass = {astro-ph.CO},
       adsurl = {https://ui.adsabs.harvard.edu/abs/2009MNRAS.398L..54N}
}

@ARTICLE{Bourne2016,
       author = {{Bourne}, Martin A. and {Power}, Chris},
        title = "{Simulating feedback from nuclear clusters: the impact of multiple sources}",
      journal = {\mnras},
         year = 2016,
        month = feb,
       volume = {456},
       number = {1},
        pages = {L20-L24},
          doi = {10.1093/mnrasl/slv162},
archivePrefix = {arXiv},
       eprint = {1510.05697},
 primaryClass = {astro-ph.GA},
       adsurl = {https://ui.adsabs.harvard.edu/abs/2016MNRAS.456L..20B}
}

@ARTICLE{Power2011,
       author = {{Power}, C. and {Zubovas}, K. and {Nayakshin}, S. and {King}, A.~R.},
        title = "{Self-regulated star formation and the black hole-galaxy bulge relation}",
      journal = {\mnras},
         year = 2011,
        month = may,
       volume = {413},
       number = {1},
        pages = {L110-L113},
          doi = {10.1111/j.1745-3933.2011.01048.x},
archivePrefix = {arXiv},
       eprint = {1103.1702},
 primaryClass = {astro-ph.CO},
       adsurl = {https://ui.adsabs.harvard.edu/abs/2011MNRAS.413L.110P}
}

@ARTICLE{Moster2018,
       author = {{Moster}, Benjamin P. and {Naab}, Thorsten and {White}, Simon D.~M.},
        title = "{EMERGE - an empirical model for the formation of galaxies since z {\ensuremath{\sim}} 10}",
      journal = {\mnras},
         year = 2018,
        month = jun,
       volume = {477},
       number = {2},
        pages = {1822-1852},
          doi = {10.1093/mnras/sty655},
archivePrefix = {arXiv},
       eprint = {1705.05373},
 primaryClass = {astro-ph.GA},
       adsurl = {https://ui.adsabs.harvard.edu/abs/2018MNRAS.477.1822M}
}

@ARTICLE{Sugimura2024,
       author = {{Sugimura}, Kazuyuki and {Ricotti}, Massimo and {Park}, Jongwon and {Garcia}, Fred Angelo Batan and {Yajima}, Hidenobu},
        title = "{Violent starbursts and quiescence induced by FUV radiation feedback in metal-poor galaxies at high-redshift}",
      journal = {arXiv e-prints},
         year = 2024,
        month = mar,
          eid = {arXiv:2403.04824},
        pages = {arXiv:2403.04824},
          doi = {10.48550/arXiv.2403.04824},
archivePrefix = {arXiv},
       eprint = {2403.04824},
 primaryClass = {astro-ph.GA},
       adsurl = {https://ui.adsabs.harvard.edu/abs/2024arXiv240304824S}
}

@ARTICLE{Jecmen2023,
       author = {{Jecmen}, Michelle C. and {Oey}, M.~S.},
        title = "{Delayed Massive-star Mechanical Feedback at Low Metallicity}",
      journal = {\apj},
         year = 2023,
        month = dec,
       volume = {958},
       number = {2},
          eid = {149},
        pages = {149},
          doi = {10.3847/1538-4357/ad0460},
archivePrefix = {arXiv},
       eprint = {2310.10589},
 primaryClass = {astro-ph.GA},
       adsurl = {https://ui.adsabs.harvard.edu/abs/2023ApJ...958..149J}
}

@BOOK{Lamers1999,
       author = {{Lamers}, Henny J.~G.~L.~M. and {Cassinelli}, Joseph P.},
        title = "{Introduction to Stellar Winds}",
         year = 1999,
       adsurl = {https://ui.adsabs.harvard.edu/abs/1999isw..book.....L}
}

@ARTICLE{Pallottini2023A&A,
       author = {{Pallottini}, A. and {Ferrara}, A.},
        title = "{Stochastic star formation in early galaxies: Implications for the James Webb Space Telescope}",
      journal = {\aap},
         year = 2023,
        month = sep,
       volume = {677},
          eid = {L4},
        pages = {L4},
          doi = {10.1051/0004-6361/202347384},
archivePrefix = {arXiv},
       eprint = {2307.03219},
 primaryClass = {astro-ph.GA},
       adsurl = {https://ui.adsabs.harvard.edu/abs/2023A&A...677L...4P}
}

@ARTICLE{Shen2023,
       author = {{Shen}, Xuejian and {Vogelsberger}, Mark and {Boylan-Kolchin}, Michael and {Tacchella}, Sandro and {Kannan}, Rahul},
        title = "{The impact of UV variability on the abundance of bright galaxies at z {\ensuremath{\geq}} 9}",
      journal = {\mnras},
         year = 2023,
        month = nov,
       volume = {525},
       number = {3},
        pages = {3254-3261},
          doi = {10.1093/mnras/stad2508},
archivePrefix = {arXiv},
       eprint = {2305.05679},
 primaryClass = {astro-ph.GA},
       adsurl = {https://ui.adsabs.harvard.edu/abs/2023MNRAS.525.3254S}
}

@ARTICLE{Sun2023ApJ,
       author = {{Sun}, Guochao and {Faucher-Gigu{\`e}re}, Claude-Andr{\'e} and {Hayward}, Christopher C. and {Shen}, Xuejian and {Wetzel}, Andrew and {Cochrane}, Rachel K.},
        title = "{Bursty Star Formation Naturally Explains the Abundance of Bright Galaxies at Cosmic Dawn}",
      journal = {\apjl},
         year = 2023,
        month = oct,
       volume = {955},
       number = {2},
          eid = {L35},
        pages = {L35},
          doi = {10.3847/2041-8213/acf85a},
archivePrefix = {arXiv},
       eprint = {2307.15305},
 primaryClass = {astro-ph.GA},
       adsurl = {https://ui.adsabs.harvard.edu/abs/2023ApJ...955L..35S}
}
\end{document}